\documentclass[conference]{IEEEtran}
\IEEEoverridecommandlockouts
\usepackage{cite}
\usepackage{amsmath,amssymb,amsfonts}
\usepackage{algorithmic}
\usepackage{graphicx}
\usepackage{textcomp}
\usepackage{xcolor}
\usepackage{amsmath}
\usepackage{enumitem}
\usepackage{multirow}
\usepackage{bm}
\usepackage{mathrsfs}
\usepackage{tcolorbox}
\usepackage{subfig}
\usepackage{booktabs}
\usepackage{colortbl}
\usepackage{url}
\def\BibTeX{{\rm B\kern-.05em{\sc i\kern-.025em b}\kern-.08em
    T\kern-.1667em\lower.7ex\hbox{E}\kern-.125emX}}
\begin{document}

\title{Practitioners' Expectations on Code Completion
}
\newcommand{\revise}[1]{\textcolor{red}{{#1}}}
\newcommand\etal{{\it{et al.\ }}}
\newcommand{\yun}[1]{\textcolor{blue}{{#1}}}
\newcommand{\wcz}[1]{\textcolor{violet}{{#1}}}

\newcommand{\finding}[2]{
\begin{tcolorbox}[width=\linewidth,boxrule=0pt,top=1pt, bottom=1pt, left=1pt,right=1pt, colback=gray!20,colframe=gray!20]
\textbf{Finding #1:} 
{#2}
\end{tcolorbox}}
\author{\IEEEauthorblockN{Chaozheng Wang}
\IEEEauthorblockA{
\textit{Harbin Institute of Technology,}\\
Shenzhen, China \\
wangchaozheng@stu.hit.edu.cn}
\and
\IEEEauthorblockN{Junhao Hu}
\IEEEauthorblockA{
\textit{Peking University,}\\
Beijing, China \\
junhaohu@stu.pku.edu.cn}
\and
\IEEEauthorblockN{Cuiyun Gao}
\IEEEauthorblockA{
\textit{Harbin Institute of Technology,}\\
Shenzhen, China \\
gaocuiyun@hit.edu.cn}
\and
\IEEEauthorblockN{Yu Jin}
\IEEEauthorblockA{
\textit{Tencent Inc.,}\\
Guangzhou, China \\
lenajin@tencent.com}
\and
\IEEEauthorblockN{Tao Xie}
\IEEEauthorblockA{
\textit{Peking University,}\\
Beijing, China \\
taoxie@pku.edu.cn}
\and
\IEEEauthorblockN{Hailiang Huang}
\IEEEauthorblockA{
\textit{Tencent Inc.,}\\
Guangzhou, China \\
eraserhuang@tencent.com}
\and
\IEEEauthorblockN{Zhenyu Lei}
\IEEEauthorblockA{
\textit{Tencent Inc.,}\\
Guangzhou, China \\
rainylei@tencent.com}
\and
\IEEEauthorblockN{Yuetang Deng}
\IEEEauthorblockA{
\textit{Tencent Inc.,}\\
Guangzhou, China \\
yuetangdeng@tencent.com}
}

\maketitle

\begin{abstract}
Code completion has become a common practice for programmers during their daily programming activities. It aims at automatically predicting the next tokens or lines that the programmers tend to use.
A good code completion tool can substantially save keystrokes and improve the programming efficiency for programmers. Recently, various techniques for code completion have been proposed for usage in practice. However, it is still unclear what are practitioners' expectations on code completion and whether existing research has met their demands. To fill the gap, we perform an empirical study by first interviewing 15 practitioners and then surveying 599 practitioners from 18 IT companies about their expectations on code completion. We then compare the practitioners' demands with current research via conducting a literature review of papers on code completion published in premier publication venues from 2012 to 2022. Based on the comparison, we highlight the directions desirable for researchers to invest efforts towards developing code completion techniques for meeting practitioners' expectations.
 
\end{abstract}

\begin{IEEEkeywords}
code completion, empirical study, practitioners' expectations
\end{IEEEkeywords}

\section{Introduction}
Code completion aims to predict the code tokens or lines that programmers tend to input in their daily programming activities. It can substantially save
keystrokes and improve coding efficiency for programmers. As reported by \cite{DBLP:conf/icsm/LiHLYT21}, with code completion, programmers can averagely save 3.65 keystrokes for completing a token.
Code completion has become the most frequently-used feature of modern integrated development environments (IDEs) \cite{DBLP:conf/wcre/AmannPNM16}, and arouse increasingly more attention from both academia and industry \cite{DBLP:conf/kdd/SvyatkovskiyZFS19,DBLP:conf/aaai/WangL21a,DBLP:conf/icse/KimZT021,DBLP:conf/iwpc/LiuLW0FJ20,DBLP:conf/msr/SvyatkovskiyLHR21,DBLP:conf/sigsoft/SvyatkovskiyDFS20,guo2021learning, DBLP:conf/icse/IzadiGG22, copilot, aixcoder, tabnine}.

Traditional code completion techniques \cite{clangd,DBLP:conf/pldi/MandelinXBK05, DBLP:conf/pldi/GveroKKP13} adopt static analysis and provide completion candidates according to pre-defined rules, which generally require non-trivial manual efforts and perform ineffectively. To automate the whole process, machine learning (ML)-based methods \cite{DBLP:conf/icse/HindleBSGD12, DBLP:conf/se/ProkschLM16, DBLP:conf/icse/NguyenN15} such as N-gram models are proposed to learn the code statistics, and thereby can generalize to new code.
With the development of deep learning (DL), 
many DL-based approaches \cite{DBLP:conf/kbse/NguyenNLW19, DBLP:conf/kdd/SvyatkovskiyZFS19} have been proposed to implicitly learn the patterns in code and achieve state-of-the-art performance.
Some of the DL-based techniques \cite{DBLP:conf/icse/KimZT021, DBLP:conf/sigsoft/SvyatkovskiyDFS20, DBLP:conf/icse/IzadiGG22, DBLP:conf/msr/CiniselliCPPPB21} employ pre-trained language (PLMs) such as BERT and GPT \cite{DBLP:conf/naacl/DevlinCLT19, brown2020language, radford2019language} for code completion, since PLMs are trained on large unlabelled corpora and can encode amount of code knowledge into large-scale parameters. Besides accurately predicting the next tokens, PLM-based methods can also perform well in predicting the next lines of code.

Despite numerous research efforts on code completion, unfortunately, no prior studies have investigated practitioners' expectations on the techniques.
It is unclear whether practitioners appreciate the current code completion techniques and what aspects (e.g., completion efficiency and effectiveness) they care most about during adoption. The practitioners' perspective is important to unveil critical problems and can provide guidance for software engineer researchers to create solutions that satisfy programmers.

In this paper, we follow a mixed-methods approach \cite{DBLP:conf/icse/HuX0WCZ22,DBLP:conf/issta/KochharXLL16} to gain insights into practitioners' expectations on code completion. We start with semi-structured interviews with 15 professionals who have an average of 8.4 years of software programming experience.
Through the interviews, we qualitatively investigate the state of code completion practices, issues faced by our interviewees when using code completion tools, and their expectations on code completion. We then perform an exploratory survey with 599 professionals from 18 IT companies to quantitatively validate practitioners' expectations uncovered in our interviews. We finally conduct a literature review of research papers published in premier venues from 2012 to 2022 and compare the techniques proposed in the papers against the expectations of practitioners.

Specifically, we investigate the following four research questions:

\noindent\textbf{RQ1: What is the state of code completion practices?}

This research question studies code completion practices
, including code completion tools and
usage scenarios (the scenarios practitioners use code completion). For completion tools, 72\% and 54\% participants express that they often use code completion in daily programming and are eager for a better code completion tool,
respectively. Meanwhile, built-in tools in IDEs are far more popular than third-party plug-in completion tools such as Copilot \cite{copilot} and IntelliCode \cite{DBLP:conf/sigsoft/SvyatkovskiyDFS20}, used by 96\% and about 13\% of the participants, respectively.

For usage scenarios, to facilitate the investigation, we broadly divide code completion techniques according to the completion granularities, including token level and statement level. On average, 81\% participants adopt token-level completion, which is evidently more than those adopting statement-level completion (only 32\%).
For token-level completion, 84\% and 85\% participants often utilize tools to complete identifiers and recommend APIs, respectively.
For statement-level completion, completing the currently edited line and predicting the API argument are the two most popular usage scenarios.

\noindent\textbf{RQ2: Is code completion important for practitioners, and what are the issues? }

In this research question, we investigate how practitioners perceive the importance of code completion, and study the issues faced by practitioners during programming. For the importance perceived by the practitioners, 87\% of them
strongly agree or agree with the importance of code completion during software development.
For the issues of code completion, only 36\% of the participants think current code completion tools are satisfying. 
56\% and 58\% participants consider \textit{erroneous completion} and \textit{painful tool installation} as the main issues, respectively.

\noindent\textbf{RQ3: What are practitioners' expectations on code completion tools? }

This research question focuses on practitioners' expectations on different granularity level code completion tools, including token-level and statement-level completion.
We investigate the expectations from multiple aspects including \textit{usage scenarios}, \textit{evaluation metrics}, \textit{access to service} (online or offline), completion \textit{effectiveness}, and \textit{efficiency}.
On average, 88\% and 57\% participants expect token-level completion scenarios and statement-level completion scenarios, respectively.
For token-level completion, about 90\% participants expect tools
to complete identifiers, recommend APIs, and complete paths (i.e., when the required tokens are related to a file/directory path). 79\% of participants are satisfied if the completion tools provide appropriate tokens within the top three candidates. Most participants (85\%) expect tools to generate completion results within 200 milliseconds. For statement-level completion, the most expected scenarios are \textit{skeleton completion} (predicting the skeleton of classes and methods), \textit{API argument recommendation} (recommending the arguments of called APIs) and \textit{completion of currently edited line}. The most favored completion latency is no more than 2 seconds. For evaluation metrics, most participants care about the completion accuracy (79\%) and grammatical correctness (86\%) for adopting token-level and statement-level completion techniques, respectively.

\noindent\textbf{RQ4: How well current research satisfies practitioners' demands? }

This research question investigates the 
code completion research and explores the gap between the research and practitioners' expectations. We have identified 26 papers about code completion techniques, among which 17 papers propose token-level completion and 10 papers are about statement-level completion\footnote{The method proposed in \cite{DBLP:conf/icse/IzadiGG22} can conduct both token-level and statement-level completion.}. For the research about token-level completion, most of them focus on \textit{API recommendation} and \textit{identifier completion} scenarios, which is consistent with practitioners' expectations. However, no papers have covered the path completion scenario, a scenario expected by the large majority of practitioners. For the research about statement-level completion, most papers focus on next line completion, but only a few papers explore \textit{skeleton completion} and \textit{API argument prediction} that practitioners expect most.
Besides, most papers measure \textit{overlapped n-grams} and \textit{edit similarity} between completed code and human-written code, however,
they are not preferred by the majority of the participants.

Our research aims at providing the future research direction on code completion and helping  researchers to consider the expectations of practitioners when studying code completion techniques. We hope researchers to propose better code completion tools, which can be eventually adapted into programming practices and satisfy a wide range of programmers.

In summary, we make the following contributions:
\begin{itemize}
    \item We are the first to shed light on the practitioners' expectations on code completion tools. We first interview 15 and then survey 599 practitioners from 18 IT companies. We investigate their opinions on code completion tools, the scenarios they use for code completion, and their expectations on code completion tools.
    \item We comprehensively review the papers published in the premier venues in software engineering and artificial intelligence fields in the last ten years. Then we compare the published papers with the practitioners' expectations, highlighting the aspects desirable for researchers to be improved in code completion for meeting the demands of practitioners.
\end{itemize}

\textbf{Paper Structure:} Section \ref{sec:method} describes the methodology of our study. Section \ref{sec:result} presents the results of our study. We discuss the implications of our results in Section \ref{sec:discuss}. Section \ref{sec:related} discusses related work. 
\section{Research Methodology}\label{sec:method}
In this section, we introduce our overall research methodology. Our research consists of three main stages. \textbf{Stage 1:}
Interviewing with professionals about their practices on code completion,
the issues they have met, and their expectations on code completion tools. \textbf{Stage 2:} Performing an online survey to validate and extend practitioners' expectations based on the interview.
\textbf{Stage 3:} Conducting
a literature review 
to analyze whether and to what extent current techniques have satisfied practitioners'
demands. The interviews and survey were approved by the relevant institutional review board (IRB).

\subsection{Stage 1: Interview}

\subsubsection{Protocol}
Two authors conduct a series of in-person, semi-structured, and in-depth interviews based on an interview guide to explore the participants' practices, issues, and expectations on code completion. We invite 15 programming professionals from seven IT companies worldwide to participate in the interviews. Each interview lasts 45-60 minutes. 

In each interview,
we first ask the interviewees some demographic questions about their background including job roles and working
experience. Then, we ask the interviewees to freely talk about what they regard as a good code completion tool and their expectations of code completion techniques.  In the end, we investigate their code completion practices and the issues they have ever met.


\subsubsection{Interviewees}
We invite interviewees from our
networks in the IT industry. The interviewees are working full-time in different roles such as developers and AI algorithms designers. Eventually, 15 interviewees from seven IT companies agree to participate in the interview. The programming experience of our interviewees varies from 5 years to 13 years, with
the average programming experience at 8.4 years.

\subsubsection{Data analysis}
The first author transcribes the interviews and summarizes the interviewees' perspectives into opinion cards. Then another author verifies the summarized
opinion cards and provides suggestions for improvement. After incorporating the suggestions, the two authors separately analyze and sort the opinion cards into potential
descriptions for the questionnaire. {The Cohen's Kappa value between the two authors is 0.67, indicating a substantial agreement between them}. The two authors discuss their disagreements to reach a common decision. To reduce the bias of the two authors in sorting descriptions,
another two authors have also reviewed and confirmed the final set of survey descriptions. 

At the end of the interview, we derive four code completion practices and six
issues. We summarize three and six usage scenarios for practitioners' expectations on token-level and statement-level code completion, respectively.
Furthermore, we derive eight and thirteen 
factors which specifically affect the adoption of token-level and statement-level completion, respectively, and draw another three aspects of general code completion tools.

\begin{figure}[t]
    \centering
    \includegraphics[width=0.47\textwidth]{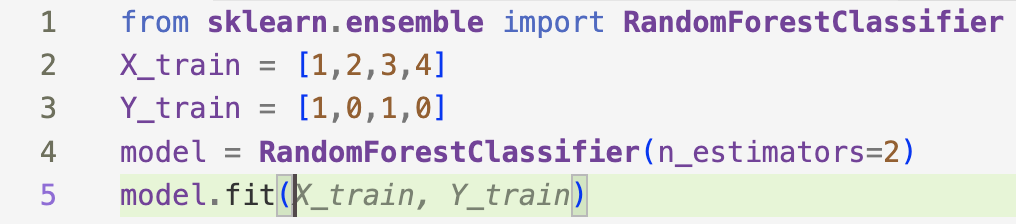}
    \caption{Example for illustrating the \textit{API argument prediction} scenario
    in the survey.}
    \label{fig:example}
\end{figure}

\subsection{Stage 2: Online Survey}\label{sec:survey}

\subsubsection{Survey Design} The survey consists of different types of questions including single/multiple-choice, Likert scale (Options: Strongly Disagree, Disagree, Neutral, Agree, Strongly Agree), and short answer questions. For each question, we add an ``I don't understand" choice to filter the cases that participants do not understand our descriptions.

The survey consists of six parts:

\begin{itemize}
    \item \textbf{Demographics.} The survey first asks for demographic information of
    the participants, including primary job roles and programming experience.
    
    \item \textbf{Code completion tools.} We ask the participants about the factors about code completion tools. In particular, we investigate the code completion tools they have ever used and their practices on using code completion tools. 
    \item \textbf{Code completion usage scenarios.} This part aims at investigating in what scenarios practitioners use code completion tools. We summarize three scenarios for token-level completion and six scenarios for statement-level completion. For each scenario, we utilize a screenshot or GIF for illustration, with an example shown in Figure \ref{fig:example}.
    The three token-level completion scenarios include \textit{identifier completion}, \textit{API recommendation}, and \textit{path completion} (i.e.,
    completing the path of a referred file or directory). The studied statement-level scenarios include \textit{next line prediction}, \textit{completion of currently edited line}, \textit{API argument
    recommendation},  \textit{string completion} (e.g., completing a log),
    \textit{skeleton prediction} and \textit{block content prediction} (e.g., predicting a method block).
    \item \textbf{Tool importance.} We investigate the practitioners' attitudes towards
    the code completion tools in this part. We ask the participants about how they perceive the importance of the tools.
    \item \textbf{Code completion issues.} This part investigates the issues faced by the participants during using the code completion tools,
    such as \textit{erroneous completion}, \textit{high completion latency}, \textit{high resource consumption} and \textit{concern about data leak}.

    \item \textbf{Practitioners' expectations.} 
    In this part, we study practitioners' expectations on code completion tools of the two granularity level, i.e., token-level completion and statement-level completion. For each granularity level, we ask multiple aspects of the expectations including
    \textit{usage scenarios}, \textit{evaluation metrics}, \textit{access to service} (online or offline), completion \textit{effectiveness} and \textit{efficiency}. Besides, we also investigate general aspects that affect practitioners' likelihood to adopt code completion tools such as \textit{time consumption of tool installation }.
\end{itemize}

At the end of the survey, we allow our participants to choose to freely provide
their comments, advice, and opinions about code completion and our survey.

To check participants' perceptions about the survey length and clarity of the descriptions, we
conduct a preliminary survey with a small set of practitioners that are different from our interviewees and survey participants before large-scale distribution. Based on the received feedback, we make minor modifications to the survey and produce a final version.
We utilize widely-used questionnaire websites \cite{wj,wj2} to distribute the survey.

\subsubsection{Participant Recruitment} We contact professionals working full time in IT companies in our social networks, and
ask for their help to complete and disperse the survey. In particular, we send invitations to the professionals working in Microsoft, Intel, Tencent, Alibaba, and other IT companies. 
We finally receive 611 survey responses in total with
the average completion time at 8.2 minutes. Among them, we discard 12 responses that are completed within
two minutes. The survey results presented in this paper are analyzed from the remaining 599 valid responses. An overview of the surveyed participants' roles and their experience is depicted in Table \ref{tab:demo}. Most participants are engaged in software development and have 3-5 years of programming
experience.

\begin{table}[]
    \centering
    \caption{Participants roles \& working experience}
    \begin{tabular}{ll|lllll}
    \toprule
     Role   & Population & $<$1y & 1-3y & 3-5y & 5-10y & $>$10y \\
     \midrule
     Development& 345&  32 & 102 & 121 & 65 & 15 \\
     Algorithm Design& 173& 10& 46& 81& 42 & 4\\
    Testing & 27& 6& 8& 4& 6& 3\\
    Architect & 4& 0& 1& 2& 1& 0 \\
    Project Manager & 2& 0& 0& 0& 1& 1 \\
    
    Others & 48 & 9& 15& 17& 7& 0 \\
    \midrule
    & 599& 57& 172& 225& 122& 23\\
    \bottomrule
    \end{tabular}
    
    \label{tab:demo}
\end{table}

\subsubsection{Result Analysis} We analyze the results based on the question types. For multiple-choice and single-choice questions, we report the percentage of each selected option. For Likert-scale questions, we 
draw bar charts
to illustrate the distributions of the Likert scores.
For the open-ended short answer questions, we conduct a qualitative analysis of the results. Besides, we drop ``I don’t understand'' ratings that form a small minority (less than 1\%) of all the received ratings.

\subsection{Stage 3: Literature Review}
The papers about code completion are usually published
in software engineering and artificial intelligence fields. Therefore, we go through research papers published in ICSE, ESEC/FSE, ASE, ICPC, SANER, MSR, ICSME, PLDI, OOSPLA, TSE, TOSEM, EMSE, ACL, EMNLP, NAACL, IJCAI, ICLR, NeurIPS, and AAAI from 2012 to 2022. We select papers from the above conferences and journals because they are premier publication venues in software engineering and artificial intelligence fields.

We first read the title and abstract of the papers to check whether they are related to code completion.
For each code completion paper, two authors read its content and analyze the capabilities of the proposed approach in terms of completion granularity, scenarios, evaluation metrics, and the access to the completion service, and efficiency. For instance, Izadi et al. \cite{DBLP:conf/icse/IzadiGG22} declare that they utilize multi-task learning to train the model to predict next token and next statement. Thus, we infer that this paper works on both token-level and statement-level completion. Svyatkovskiy et al. \cite{DBLP:conf/kdd/SvyatkovskiyZFS19} claim to provide web completion services and client completion model, and we infer the access to its service to be both online and offline. Two authors discuss the differences in the capability analysis and confirm the final results through further paper reading.
\section{Result Analysis}\label{sec:result}

\subsection{RQ1: Code completion practices}\label{sec:prac}
\begin{figure}[t]
    \centering
    \subfloat[Statistics of code completion tools that practitioners have used.]{\includegraphics[width=0.47\textwidth]{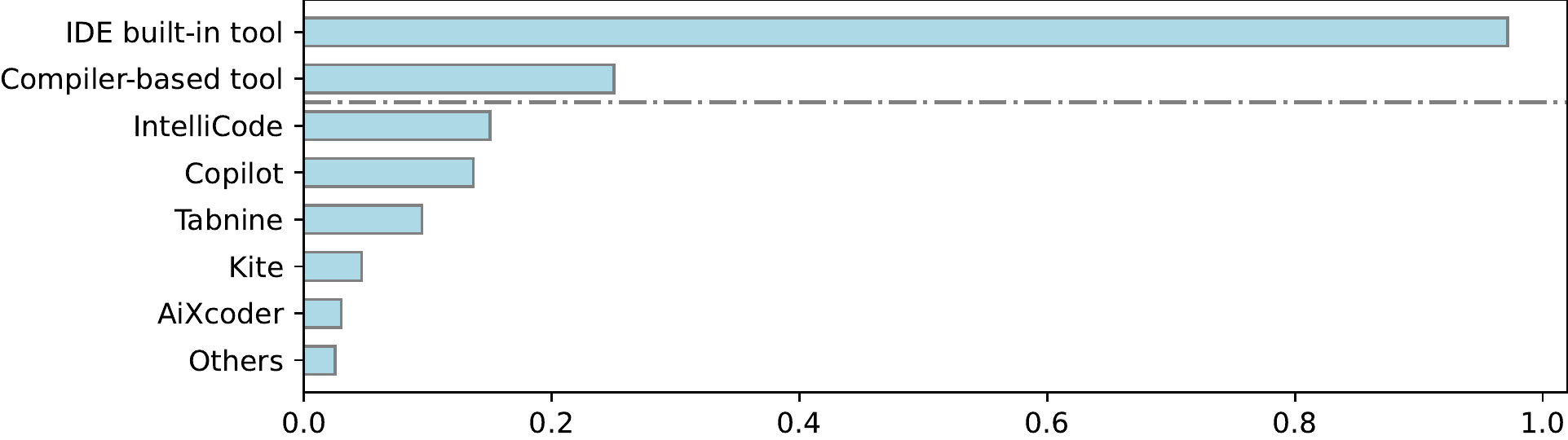}\label{fig:tool}}
    
    \subfloat[Usage status of code completion tools.
    ]{\includegraphics[width=0.47\textwidth]{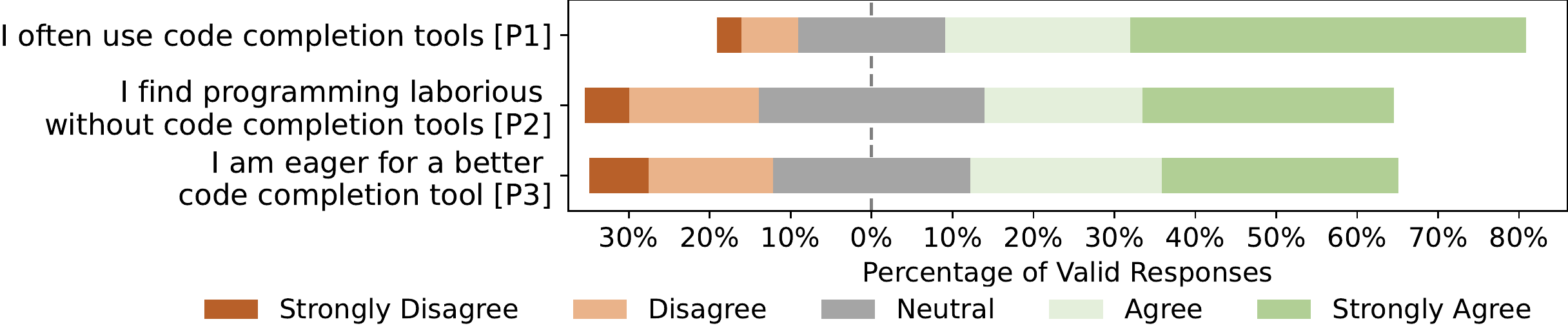}\label{fig:practices}}
    \caption{Code completion tools.}
    \label{fig:prac}
\end{figure}

\begin{figure}[t]
    \centering
    \includegraphics[width=0.47\textwidth]{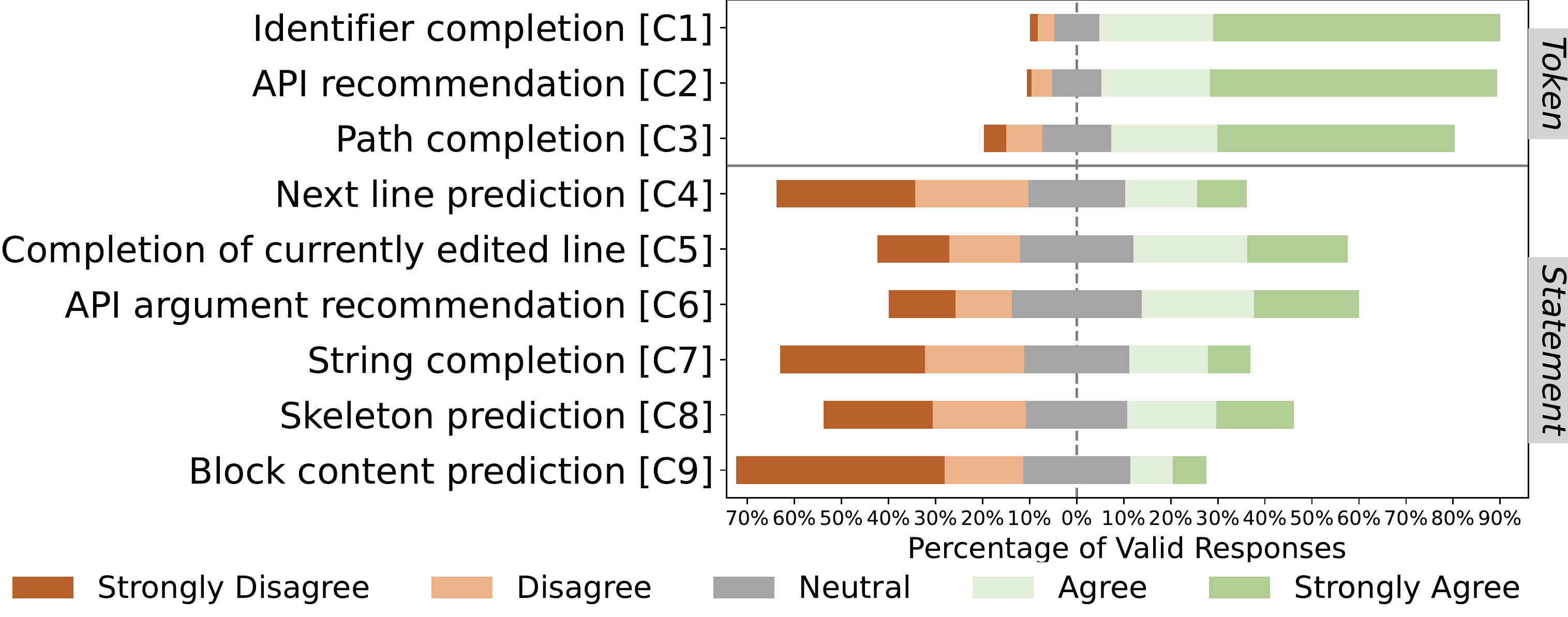}
    \caption{Scenarios that practitioners use code completion.}
    \label{fig:contexts}
\end{figure}

In this research question, we investigate the practitioners' code completion practices, including the used
code completion tools and
usage scenarios. The participants' rating results for some descriptions related to code completion practices are illustrated in Figure \ref{fig:prac} and \ref{fig:contexts}.

\subsubsection{Code completion tools}

We first ask the participants what code completion tools they have used and the results are shown in Figure \ref{fig:prac} (a). From the results, we find that 96\% participants express they adopt
IDE built-in code completion tools.
Compiler-based tools such as \textit{CCLS} and \textit{Clangd} \cite{clangd} are the second most popular tools among surveyed practitioners. However, the third-party plug-in tools are not prevalent, e.g., about 13\% participants have used Copilot \cite{copilot} and IntelliCode \cite{DBLP:conf/sigsoft/SvyatkovskiyDFS20}.

We then investigate the practitioners' usage status of
code completion tools, with results illustrated
in Figure \ref{fig:prac} (b). Although most practitioners (72\%) indicate that they often adopt code completion in daily programming, only 50\% participants agree that they find programming laborious without code completion.
Besides, 54\% of them are proactive in seeking a better code completion tool.
``\textit{I always keep tuned on the release of new code completion tools and am eager to be the first to use them.}", as a participant stated.

\finding{1}{The most commonly-used code completion tool is IDE built-in tool. Other third-party plub-in tools such as IntelliCode \cite{DBLP:conf/msr/SvyatkovskiyLHR21} and Copilot \cite{copilot} are far less popular, for which only about 13\% participants have used. In addition, most
participants express that they often use code completion tools, and
54\% of them are eager for a better code completion tool.}

\subsubsection{Usage scenarios}

In this section, we investigate in what scenarios the practitioners use code completion. According to \cite{DBLP:conf/nips/LuGRHSBCDJTLZSZ21}, we study the scenarios from two granularities including token-level and statement-level completion, respectively. Token-level completion aims at completing
the currently typing token and the next token to be used for developers. Statement-level completion is able to predict multiple tokens and even lines of code. We provide the definitions of the two granularity level completion
in the survey for facilitating participants' understanding. The token-level completion and statement-level completion involve three and six scenarios, respectively, as introduced in Section \ref{sec:survey}. The participants' ratings of the usage scenarios are shown in Figure \ref{fig:contexts}. We observe that the overall popularity of statement-level completion scenarios is obviously less than that of token-level (32\% v.s. 81\% on average). 


\noindent\textbf{Token-level completion.} 
As can be seen in the upper part of Figure \ref{fig:contexts}, all the three usage scenarios are popular among the practitioners.
Specifically, about 85\% participants use code completion to complete 
identifiers and recommend APIs. 
More than 70\% survey practitioners adopt code completion for \textit{path completion} in their daily programming.
\finding{2}{All the three token-level code completion usage scenarios are popular among the participants, among which \textit{identifier completion} and \textit{API recommendation} are the most widely used scenarios. 
}

\noindent\textbf{Statement-level completion.} 
As shown in the lower part of Figure \ref{fig:contexts},
the three most popular statement-level completion scenarios
are \textit{API argument recommendation}, \textit{completion of currently edited line}, and \textit{skeleton prediction}, used by 46\%, 45\%, and 36\% participants, respectively. 
In addition, about 25\% of them adopt code completion to predict the next line of code and string content.
Only 16\% of survey participants adopt code completion in the \textit{block content prediction} scenario. One participant
said that 
``\textit{The performance of existing completion tools is not good enough
to predict the whole block content. The predicted results may be severely misleading in some cases, which reduces my programming efficiency.}"

\finding{3}{The most adopted statement-level code completion scenarios are \textit{API argument recommendation}, \textit{completion of currently edited line}, and \textit{skeleton prediction}. }


\begin{figure}[t]
    \centering
    \includegraphics[width=0.47\textwidth]{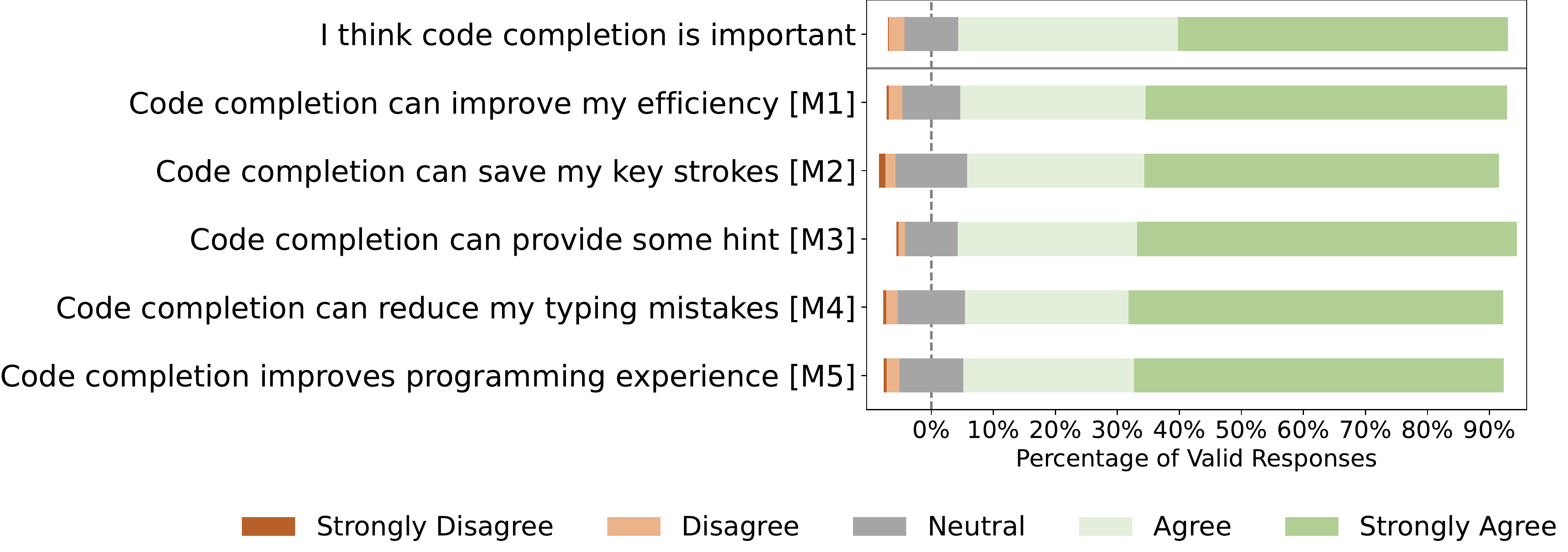}
    \caption{Importance of code completion.
    }
    \label{fig:importance}
\end{figure}

\begin{figure}[t]
    \centering
    \includegraphics[width=0.47\textwidth]{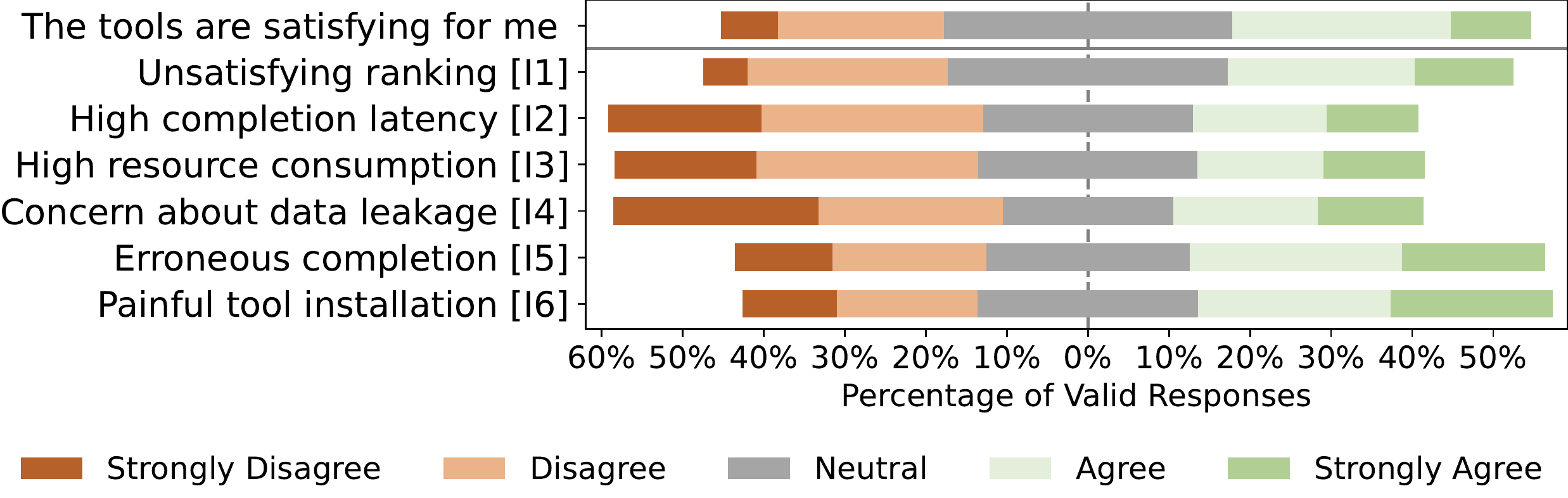}
    \caption{Issues of code completion.}
    \label{fig:issues}
\end{figure}
\subsection{RQ2: Code completion Importance and Issues}

\subsubsection{Importance of Code Completion}
We ask whether they think code completion is important for development, and their agreements on the possible reasons of the importance. The results are illustrated in Figure \ref{fig:importance}.


From the results, we can find that most participants (88\%) think that code completion plays an important role. 90\% of them agree or strongly agree that code completion can provide some hint on the implementing code. One practitioner described that
``\textit{Sometimes I forget how to type a long identifier or method name. In this case, code completion can show me the candidates and provide some hint for completing the code.}''.
Besides, more than 85\% of them think completion tools are helpful to reduce typing mistakes, improve programming efficiency, save keystrokes, and improve programming experience.
\finding{4}{88\% of the survey participants agree that code completion tools play an important role in programming. Most participants (90\%, 87\% and 86\%) think the tools can provide some hint, improve programming efficiency, and improve their programming experience.}

\subsubsection{Code completion issues}
Figure \ref{fig:issues} illustrates the participants' ratings of code completion-related issues they faced during programming.
Despite the importance of code completion, only 36\% participants think the tools they are currently using are satisfying. 45\% and 44\% participants consider \textit{erroneous completion} and \textit{painful tool installation} as the main issues, respectively. \textit{Erroneous completion} indicates that some errors occur in the completed code, which practitioners need to modify. As a participant said: ``\textit{The completion is usually not exactly what I need, and it is annoying for me to rectify the errors.}".  Besides, 34\% and 31\% of them are not satisfied with the rankings of appropriate tokens in the candidates and concerned about data leakage in the tools, respectively.
Among the participants, about a quarter of them hold the view that the high resource (i.e., memory and CPU) consumption and high completion latency
are not acceptable. A participant working
on large C++ projects told us: ``\textit{The C++ code completion tools like \textit{CCLS} and \textit{Clangd} parse the opened cpp files and build indexes for them when programming, which may consume all of my CPU and memory.}''.


\finding{5}{The most concerned
issues by practitioners are the \textit{erroneous code completion}, \textit{painful tool installation}, and \textit{unsatisfactory ranking}.
}




\subsection{RQ3: Practitioners' Expectation}
\begin{figure}
    \centering
    \subfloat[Expectations on usage scenarios.]{\includegraphics[width=0.47\textwidth]{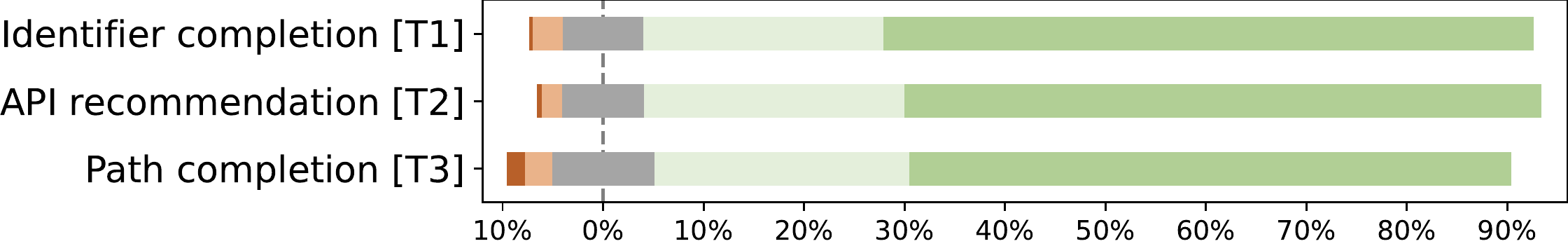}\label{fig:exp_token_context}}
    \hfill
    \subfloat[Expectations on evaluation metrics.]{\includegraphics[width=0.47\textwidth]{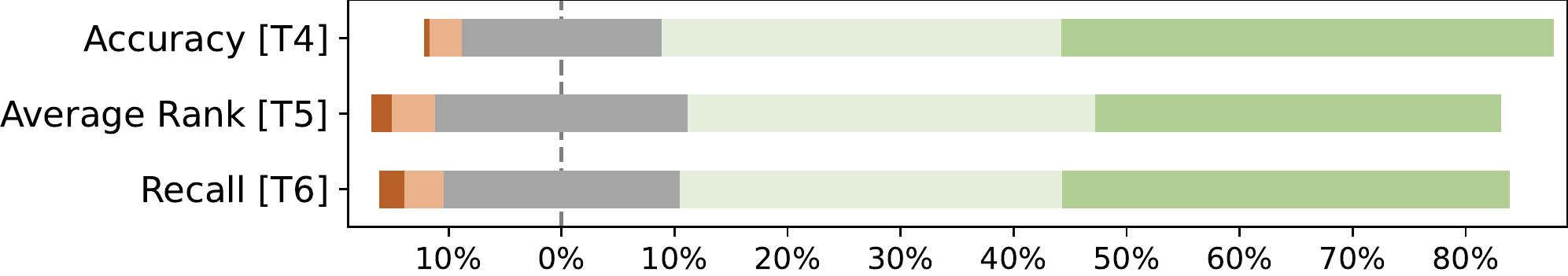}\label{fig:exp_token_metric}}
    
    \hfill
    \subfloat[Expectations on access to completion service.]{\includegraphics[width=0.47\textwidth]{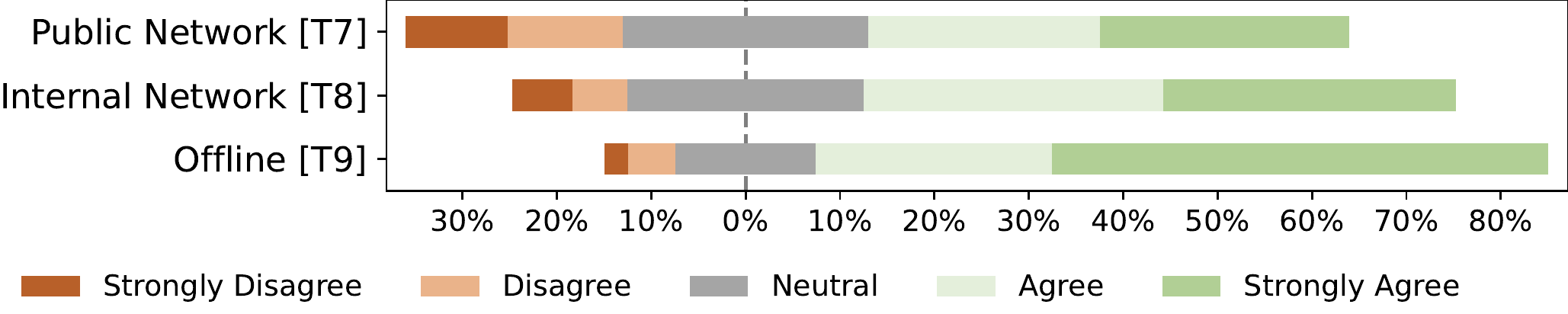}\label{fig:exp_token_access}}
    
    \hfill
    \subfloat[Expectations on code completion effectiveness.]{\includegraphics[width=0.47\textwidth]{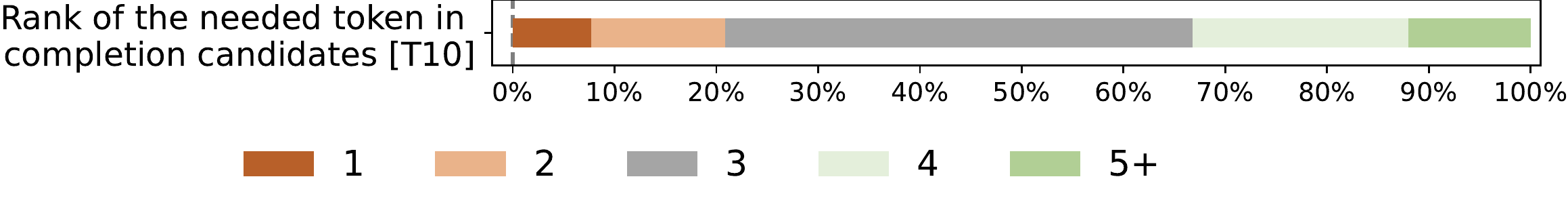}\label{fig:exp_token_effect}}
    
    \hfill
    \subfloat[Expectations on code completion efficiency.]{\includegraphics[width=0.47\textwidth]{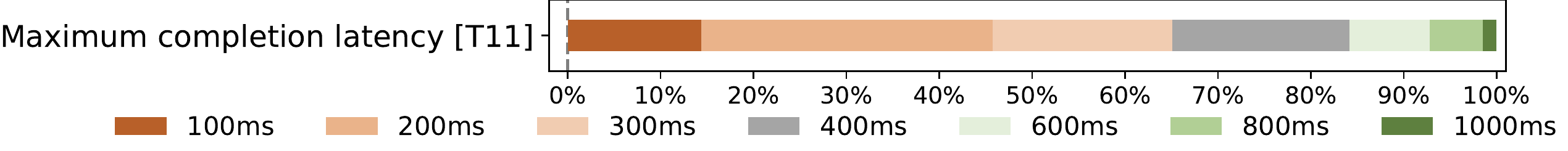}\label{fig:exp_token_time}}
    \caption{Expectations on token-level code completion, where horizontal axis denotes percentage of valid responses.}
    \label{fig:exp_token}
\end{figure}
In this research question, we comprehensively investigate the practitioners' expectations on code completion. We explore the aspects including \textit{usage scenarios}, \textit{evaluation metrics}, \textit{access to service}, completion \textit{effectiveness}, and \textit{efficiency}. We report practitioners' expectations on the two levels of granularity of code completion, i.e., token-level and statement-level completion. We also investigate practitioners' expectations on three other aspects of general completion tools that impact their usage experience, 
including \textit{time consumption of installation}, \textit{personalized completion} and \textit{additional information display}.
\subsubsection{RQ 3.1 Expectations on token-level completion}
The participants' ratings for the expectations on
token-level code completion is shown in Figure \ref{fig:exp_token}. 

\noindent\textbf{Usage scenarios.}
From Figure \ref{fig:exp_token} (a), most participants show positive attitudes towards the three scenarios. More than 88\% of them expect to use code completion in \textit{Identifier completion} and \textit{API recommendation} scenarios. Besides, 86\% of participants expect code completion tools to complete the path they are typing. In one participant's opinion, \textit{path completion} is very important when he works on a large project,
because he/she finds it very easy to make mistakes on file names without \textit{path completion}. 

\finding{6}{For token-level code completion, more than 80\% participants agree that tools are supposed to support \textit{identifier completion}, \textit{API recommendation} and \textit{path completion}.
}

\noindent\textbf{Evaluation metrics.} We study participants' opinions on evaluating token-level code completion tools in Figure \ref{fig:exp_token} (b). We observe that \textit{Accuracy} gains the most support rate (79\%). Besides, more than 70\% of participants also care about \textit{Average rank} of the needed token in the candidates (e.g., Mean Reciprocal Rank MRR) and \textit{Recall} of the tool. 

\noindent\textbf{Access to service}. Figure \ref{fig:exp_token} (c) shows that 82\% of participants expect to use token-level completion tools offline. One participant stated that: ``\textit{Offline tools are the most stable ones, which will not be affected by network status and I can use them anywhere.}". In contrast, the support rate of online tools is relatively low. 62\% and 50\% of participants are willing to use the online tools in the internal network and in the public network, respectively.

\noindent\textbf{Effectiveness.}
Figure \ref{fig:exp_token} (d) shows the practitioners' satisfaction ratio against different ranks of needed token in completion candidates. From our survey, if the code completion tool can rank the needed token in the top 3 candidates, it will satisfy about 80\% practitioners. 

\noindent\textbf{Efficiency.}
We show the participants seven GIFs with
different completion latency including 100, 200, 300, 400, 600, 800, and 1000 milliseconds. From Figure \ref{fig:exp_token} (e), we find that if a token-level code completion tool can give completion candidates within 200 milliseconds, it can satisfy 85\% practitioners. 

\subsubsection{RQ 3.2 Expectations on statement-level completion}
\begin{figure}
    \centering
    \subfloat[Expectations on usage scenarios.]{\includegraphics[width=0.47\textwidth]{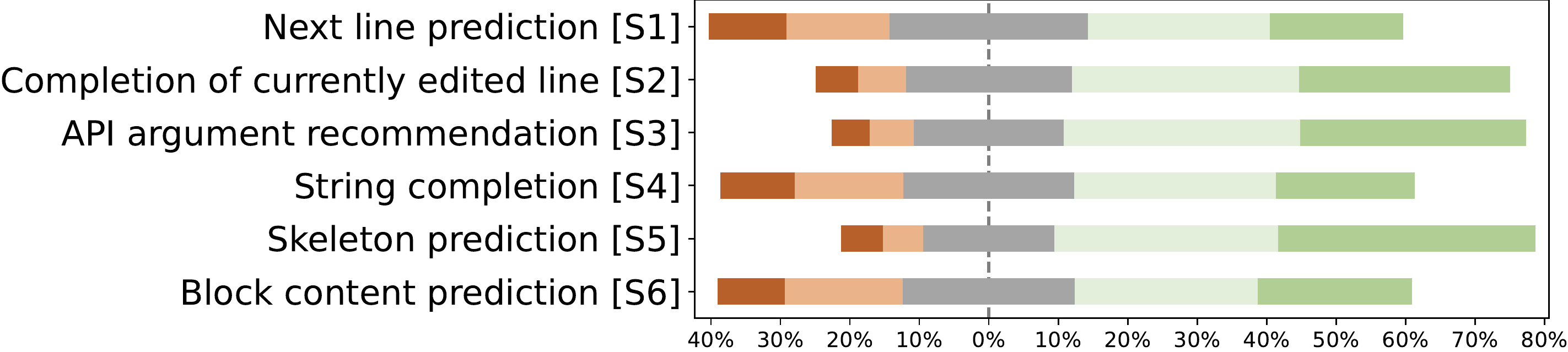}\label{fig:exp_state_context}}
    \hfill
    \subfloat[Expectations on evaluation metrics.]{\includegraphics[width=0.47\textwidth]{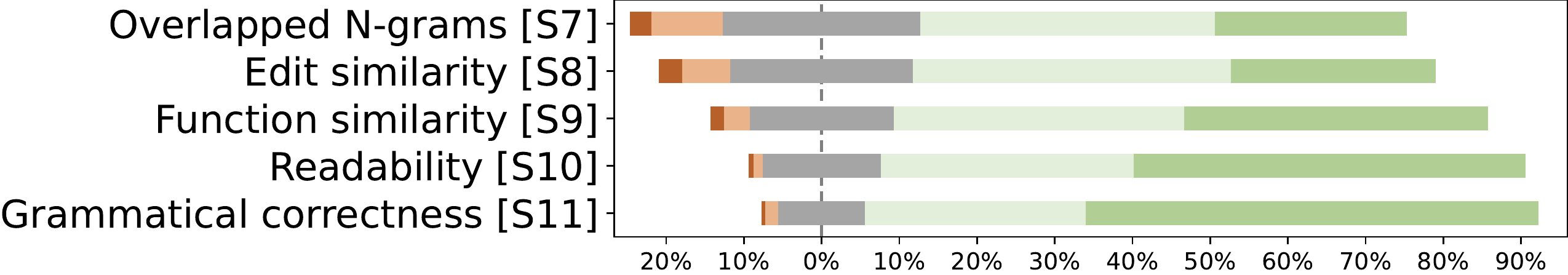}\label{fig:exp_state_metric}}
    
    \hfill
    \subfloat[Expectations on access to completion service.]{\includegraphics[width=0.47\textwidth]{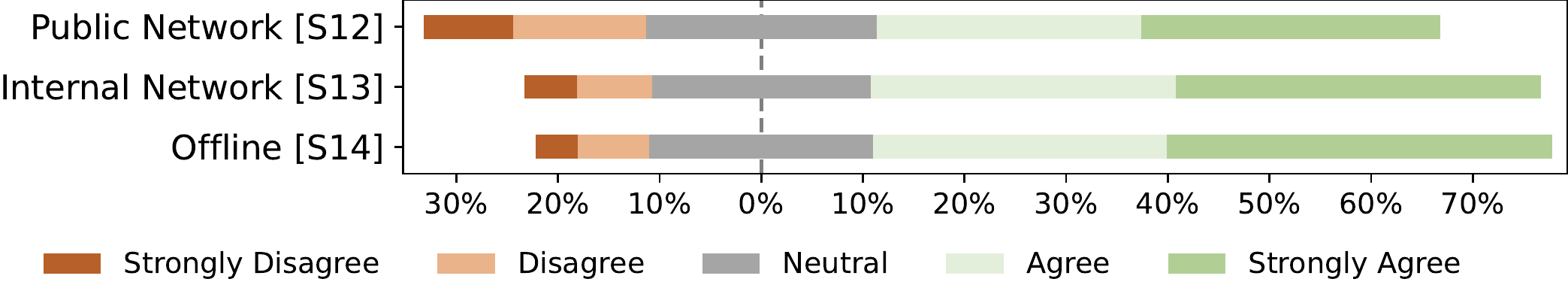}\label{fig:exp_state_access}}
    
    \hfill
        \subfloat[Expectations on code completion effectiveness.]{\includegraphics[width=0.47\textwidth]{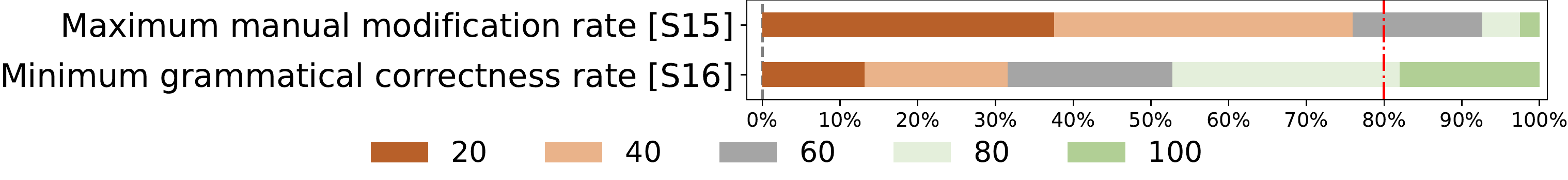}\label{fig:exp_state_effect}}
    \hfill
    \subfloat[Effectiveness v.s. Generated code quantity.]{\includegraphics[width=0.47\textwidth]{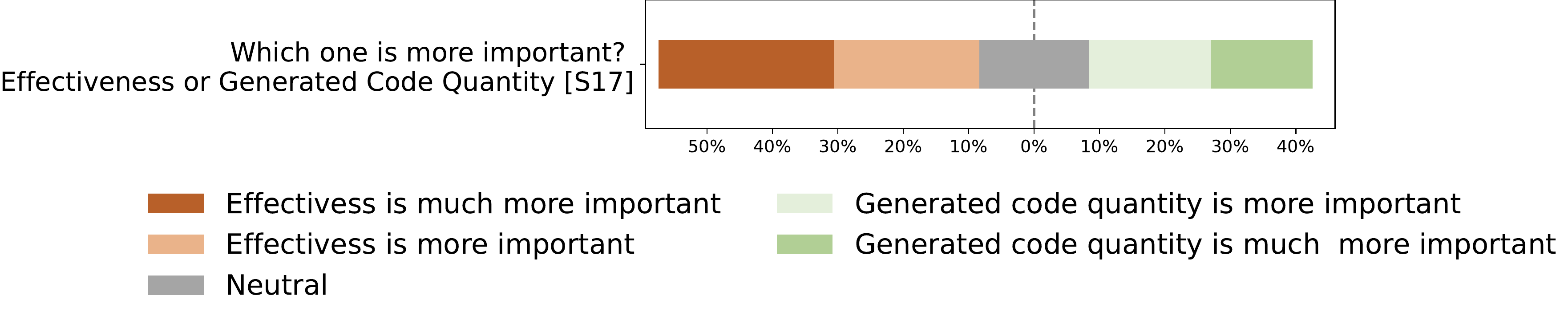}\label{fig:e_vs_q}}
    
    \hfill
    \subfloat[Expectations on average time consumption of generating a line of code.]{\includegraphics[width=0.47\textwidth]{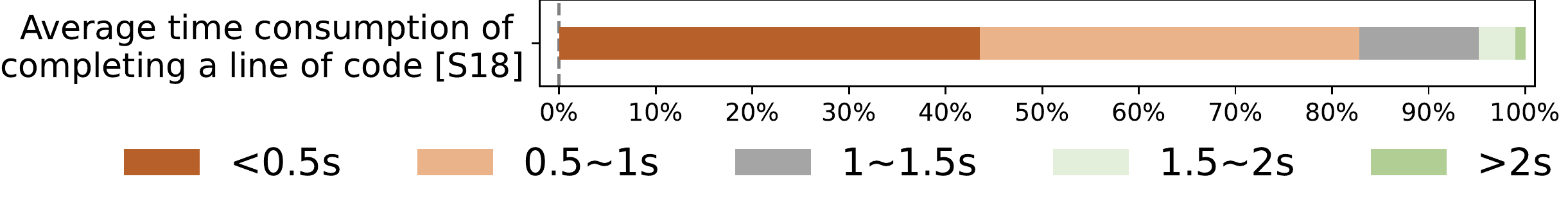}\label{fig:exp_state_time1}}
    
    \hfill
    \subfloat[Expectations on maximum time consumption of giving completion results.]{\includegraphics[width=0.47\textwidth]{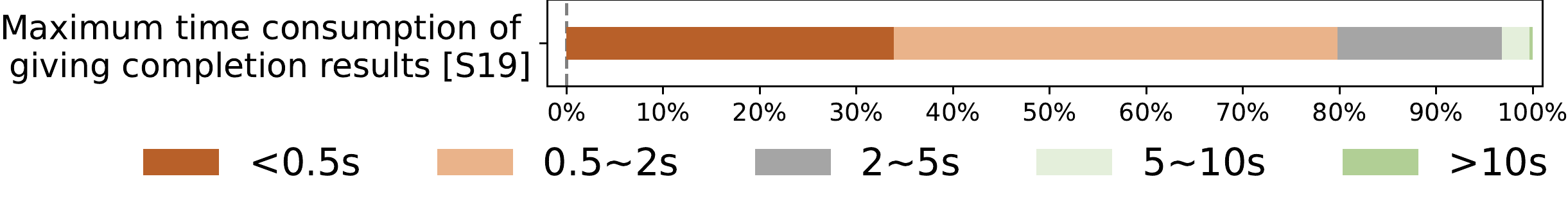}\label{fig:exp_state_time2}}
    
    \caption{Expectations on statement-level code completion, where the horizontal axis denotes percentage of valid responses.}
    \label{fig:exp_state1}
\end{figure}

The results of participants' expectations on statement-level completion can be accessed in Figure \ref{fig:exp_state1}.

\noindent\textbf{Usage scenarios.}
The practitioners' expectations on the statement-level code completion scenarios  are illustrated in Figure \ref{fig:exp_state1} (a).
We observe that practitioners' expectations show a similar trend with the popularity of usage scenarios as shown in Section \ref{sec:prac}. Likewise,
The three most expected scenarios are \textit{skeleton prediction}, \textit{completion of currently edited line} and \textit{API argument recommendation}, where all of them gain at least 60\% support rate, and \textit{skeleton prediction} is the most expected. Besides, 51\% participants agree that statement-level completion tools are supposed to predict block content such as method block and loop block. 53\% of them expect to use code completion tools to predict strings.
Interestingly, though the scenario of predicting the next line of code is kind of similar to completing the currently edited line, only 43\% of participants regard \textit{next line prediction} as an important scenario while the \textit{currently edited line completion} obtaining a 62\% support rate. 
As one participant commented: ``\textit{When finishing a line of code, I will start to type the next line immediately instead of waiting for completion results at the end of the line.}".

\finding{7}{\textit{skeleton prediction}, \textit{completion of currently edited line} and \textit{API argument recommendation} are considered to be the most expected statement-level code completion scenarios, which are similar as the popularity of the usage scenarios. 
}
    
    
    
\noindent\textbf{Evaluation metrics.} We illustrate the practitioners' expectations on different evaluation metrics for statement-level code completion approaches in Figure \ref{fig:exp_state1} (b). The top three preferred evaluation metrics are \textit{grammatical correctness}, \textit{ readability} and \textit{functional/structure similarity}, supported by more than 80\% participants.
However, the \textit{overlapping n-grams} (e.g., BLEU score \cite{DBLP:conf/acl/PapineniRWZ02}) and \textit{edit similarity} receive the least support rate (i.e., 65\% and 70\%, respectively).

\noindent\textbf{Access to service.}
Figure \ref{fig:exp_state1} (c) presents the likelihood of different access to use a statement-level code completion tool. Although the support rate of offline service
drops 15\% compared to token-level completion (from 82\% to 67\%), offline tools are still the most expected. However, the willingness to adopt an online statement-level code completion tool increases. For instance, 66\% participants agree or strongly agree to invoke online completion services in internal networks (compared with 62\% for token-level completion). Furthermore, the support rate of public-net tools also increases from 50\% to 58\%. One participant told us: ``\textit{Currently almost all statement-level completion tools are AI-based, which means that their computation complexity is high. 
Thus, utilizing online code completion service is acceptable."}.
Moreover, we observe the difference between the participants' expectations on
tools deployed in public and internal network (58\% and 66\%). According to our survey, the difference can be attributed to the concern about data leakage and network stability in public network.
As a participant stated that in his/her department, Copilot was not permitted to be installed because it would upload their code to the cloud and
might result in leakage of confidential code.

\noindent\textbf{Effectiveness.}
From Figure \ref{fig:exp_state1} (d), we find that only 26\% participants can accept the ratio of manual modification greater than 60\% of completed code. Besides, 35\% and 39\% of our participants expect the maximum modification ratio to be no more than 20\% and 40\%. For grammatical correctness, if a statement-level code completion tool ensures that at least 80\% of the completed code is grammatically correct, it can satisfy 82\% practitioners. Besides, 18\% participants have strict requirements on the correctness of generated code, i.e., they expect the tools to make no mistake on code grammar.

According to existing work \cite{DBLP:conf/msr/CiniselliCPPPB21, guo2021learning}, the more code is completed, the more errors occur. We further investigate the practitioners' preference between completion effectiveness (generate less code with higher accuracy) and generated code quantity (generate more code with more manual modifications).
The results are illustrated in Figure \ref{fig:exp_state1} (e). From our survey, there is no very clear winner between them (46\% support rate for effectiveness and 40\% for generated code quantity), and
a few more participants consider effectiveness more important.
In this question, participants mainly have two points of view:

\begin{enumerate}
    \item \textbf{Effectiveness is more important.}
    For the supporters of effectiveness, completion results with errors may affect their efficiency.
    A participant shared his/her experience with
    statement-level code completion tools. ``\textit{It was very annoying for me to modify the completion results.
    ...
    I'd rather have less code completed than modify completion results.}"

    \item \textbf{Code quantity is more important.} 
    Supporters of code quantity expect statement-level completion tools to predict more code at once, even if the code needs to be modified. One participant stated that as long as the tool could predict the code with the right logic (e.g., correct code skeleton), the tool was effective in his eyes. 
\end{enumerate}


\noindent\textbf{Efficiency.} 
We study practitioners' expectations on the efficiency of statement-level code completion tools from two aspects including average time consumption of completing a line (shown in Figure \ref{fig:exp_state1} (f)) and maximum latency of generating completion results (shown in Figure \ref{fig:exp_state1} (g)), respectively.

From our survey, we observe that the speed of averagely taking 0.5 seconds to generate a line receives the highest support rate at 44\%, and 39\% participants expect the time to be no more than 1 second.
For the maximum completion latency, the practitioners who expect the code completion tools to generate completion results in less than 2 seconds have the largest proportion (i.e., 47\%). If a statement-level code completion tool generates completion results within 2 seconds, it can satisfy 67\% practitioners. In addition, few participants (i.e., 4\%) express they are willing to wait for more than 5 seconds for completion.

\finding{8}{For completion efficiency, most practitioners expect statement-level code completion tools to take less than 0.5 seconds to complete a line of code on average. Besides, to meet the majority (67\%) of practitioners' demands, tools are supposed to generate completion results within 2 seconds.}


\begin{figure}
    \centering
    \subfloat[Expectations on time consumption of tool installation.]{\includegraphics[width=0.47\textwidth]{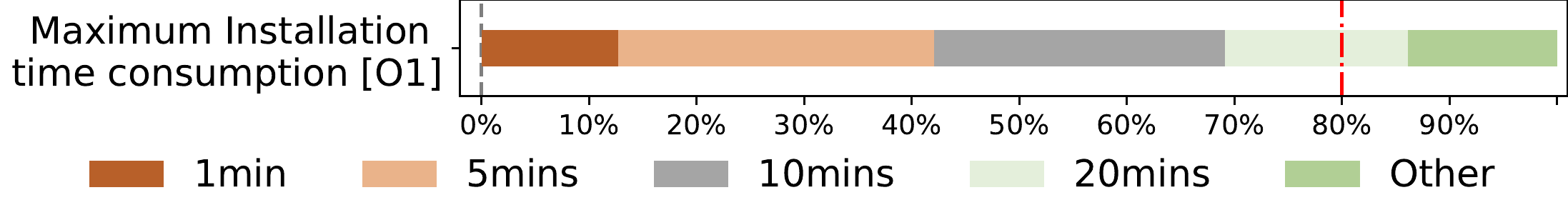}\label{fig:other_install}}
    
    \hfill
    \subfloat[Expectations on additional information display and personalized completion. ]{\includegraphics[width=0.47\textwidth]{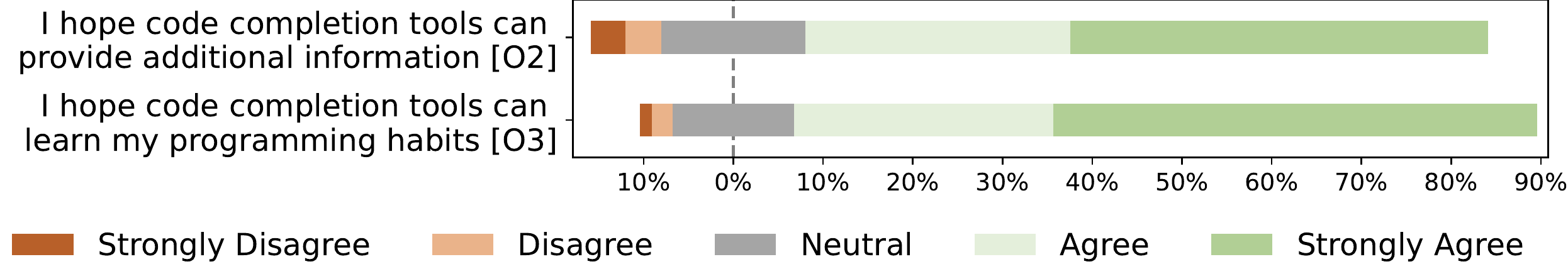}\label{fig:other_human}}
    \caption{Expectations on other aspects of general code completion tools, where the horizontal axis denotes percentage of valid responses.}
    \label{fig:exp_other}
\end{figure}

\subsubsection{RQ 3.3 Expectations on other aspects of general code completion tools} In this section we present the practitioners' expectations on general aspects of code completion, and the results are illustrated in Figure \ref{fig:exp_other}.

%

Figure \ref{fig:exp_other} (a) shows that practitioners' expectations on the time cost of installing code completion tools. 83\% participants will be satisfied if they can install the tool within 5 minutes.
Besides, we observe that 14\% of participants are willing to spend more than 20 minutes on installing tools. As one participant said: ``\textit{I would like to spend a whole day on installing a code completion tool as long as it can really improve my efficiency.}"

From Figure \ref{fig:exp_other} (b) we observe that most (76\%) practitioners agree or strongly agree that code completion tools are supposed to learn their programming habits (e.g., preference for function names
and variables). In addition, 83\% of them expect code completion tools to simultaneously present some additional information about the predicted code (e.g., showing users the API definition or documentation
while recommending an API). One participant told us: ``\textit{Sometimes I forgot the order of arguments in the API, thus showing me how the API is defined can greatly reduce my programming} mistake."

\begin{table}[!htb]\centering
\caption{Capabilities of current research. Likert score denotes the average weighted score of participants' ratings (1 to 5 correspond to strongly disagree to strongly agree).}\label{tab:SOTA}
\scriptsize
\resizebox{\linewidth}{!}{
\begin{tabular}{|l|r|c|r|l|}\hline
&\multirow{2}{*}{Description} & &Likert & \multirow{2}{*}{Papers} \\
& & &Score & \\\hline
\multirow{19}{*}{\rotatebox{90}{Token-level completion}} & \multicolumn{4}{|c|}{\cellcolor{lightgray}Usage Scenarios} \\
\cline{2-5}
& Identifier completion& [T1]& 4.50 &\cite{DBLP:conf/ijcai/LiWLK18, DBLP:conf/icml/BielikRV16, DBLP:conf/oopsla/RaychevBV16, DBLP:conf/icse/IzadiGG22, DBLP:conf/sigsoft/NguyenNNN13, DBLP:conf/kbse/LiuLZJ20, DBLP:conf/icse/KimZT021} \\
\cline{2-5}
&\multirow{3}{*}{API recommendation}& \multirow{3}{*}{[T2]}&  &\cite{DBLP:conf/ijcai/LiWLK18, DBLP:conf/sigsoft/HellendoornD17, DBLP:conf/acl/LuDHGHS22, DBLP:conf/sigsoft/TuSD14, DBLP:conf/icse/HindleBSGD12,DBLP:conf/kdd/SvyatkovskiyZFS19}\\
& & & 4.50 &\cite{DBLP:conf/kbse/LiuLZJ20, DBLP:conf/icse/KimZT021, DBLP:conf/aaai/WangL21a, DBLP:conf/iwpc/LiuLW0FJ20, DBLP:conf/icml/BielikRV16, DBLP:conf/oopsla/RaychevBV16 } \\
& & & &\cite{DBLP:conf/icse/IzadiGG22, DBLP:conf/icse/NguyenN15, DBLP:conf/sigsoft/NguyenNNN13, DBLP:conf/pldi/RaychevVY14,DBLP:conf/msr/SvyatkovskiyLHR21} \\
\cline{2-5}
&Path completion &[T3]& 4.39 & - \\
\cline{2-5}
&\multicolumn{4}{|c|}{\cellcolor{lightgray}Evaluation Metrics} \\
&Accuracy &[T4] & 4.19&\cite{DBLP:conf/icse/IzadiGG22, DBLP:conf/icml/BielikRV16,DBLP:conf/sigsoft/NguyenNNN13, DBLP:conf/icse/HindleBSGD12} \\
\cline{2-5}
&Average rank & [T5] &4.00 &\cite{DBLP:conf/sigsoft/HellendoornD17, DBLP:conf/sigsoft/TuSD14, DBLP:conf/icse/KimZT021, DBLP:conf/msr/SvyatkovskiyLHR21, DBLP:conf/kdd/SvyatkovskiyZFS19} \\
\cline{2-5}
&\multirow{3}{*}{Recall} & \multirow{3}{*}{[T6]} & &\cite{DBLP:conf/oopsla/RaychevBV16, DBLP:conf/icse/NguyenN15, DBLP:conf/sigsoft/HellendoornD17, DBLP:conf/sigsoft/NguyenNNN13} \\
& & & 4.05&\cite{ DBLP:conf/ijcai/LiWLK18, DBLP:conf/pldi/RaychevVY14, DBLP:conf/kbse/LiuLZJ20, DBLP:conf/msr/SvyatkovskiyLHR21} \\ 
& & & & \cite{DBLP:conf/iwpc/LiuLW0FJ20, DBLP:conf/kdd/SvyatkovskiyZFS19} \\\cline{2-5}
&Others & & &\cite{DBLP:conf/icml/BielikRV16, DBLP:conf/sigsoft/HellendoornD17, DBLP:conf/acl/LuDHGHS22, DBLP:conf/icse/HindleBSGD12} \\
&\multicolumn{4}{|c|}{\cellcolor{lightgray}Access to service} \\
&Online & [T7] & 3.43 &\cite{DBLP:conf/kdd/SvyatkovskiyZFS19, DBLP:conf/icse/IzadiGG22} \\
\cline{2-5}
&Offline & [T9] & 4.20 &\cite{DBLP:conf/icse/HindleBSGD12, DBLP:conf/msr/SvyatkovskiyLHR21, DBLP:conf/kdd/SvyatkovskiyZFS19} \\
&\multicolumn{4}{|c|}{\cellcolor{lightgray}Time Consumption} \\
&\multirow{2}{*}{Mentioned} & & &\cite{DBLP:conf/icml/BielikRV16, DBLP:conf/oopsla/RaychevBV16, DBLP:conf/iwpc/LiuLW0FJ20, DBLP:conf/sigsoft/TuSD14, DBLP:conf/pldi/RaychevVY14, DBLP:conf/sigsoft/NguyenNNN13} \\
& & & & \cite{DBLP:conf/sigsoft/HellendoornD17, DBLP:conf/icse/HindleBSGD12, DBLP:conf/icse/IzadiGG22, DBLP:conf/msr/SvyatkovskiyLHR21, DBLP:conf/kdd/SvyatkovskiyZFS19} \\\hline

\multirow{20}{*}{\rotatebox{90}{Statement-level completion}}&\multicolumn{4}{|c|}{\cellcolor{lightgray}Usage Scenarios} \\
&\multirow{2}{*}{Next line prediction}& \multirow{2}{*}{[S1]}& 3.27 &\cite{DBLP:conf/msr/CiniselliCPPPB21, DBLP:conf/icse/IzadiGG22, guo2021learning, DBLP:conf/sigsoft/SvyatkovskiyDFS20} \\
& & & &\cite{DBLP:conf/emnlp/ClementLLTDDSS21, DBLP:conf/acl/LuDHGHS22, DBLP:conf/kbse/YangJ0SGL17, DBLP:conf/kbse/NguyenNLW19} \\
\cline{2-5}
&Completion of currently edited line & [S2] & 3.74 &\cite{DBLP:conf/msr/CiniselliCPPPB21, DBLP:conf/sigsoft/SvyatkovskiyDFS20, DBLP:conf/kbse/NguyenNLW19} \\
\cline{2-5}
&API argument recommendation &[S3] & 3.81 &\cite{DBLP:conf/msr/CiniselliCPPPB21, DBLP:conf/sigsoft/SvyatkovskiyDFS20, DBLP:conf/sigsoft/NguyenNNN13} \\
\cline{2-5}
\cline{2-5}
&String completion &[S4] & 3.32& - \\ 
\cline{2-5}
&Skeleton prediction &[S5] & 3.88 & - \\ 
\cline{2-5}
&Block content prediction & [S6] & 3.43 &\cite{DBLP:conf/msr/CiniselliCPPPB21, DBLP:conf/emnlp/ClementLLTDDSS21, DBLP:conf/icse/WenA0LB21} \\\cline{2-5}
&\multicolumn{4}{|c|}{\cellcolor{lightgray}Evaluation Metrics} \\
&Overlapped n-grams &[S7]&3.72 &\cite{DBLP:conf/emnlp/ClementLLTDDSS21, DBLP:conf/acl/LuDHGHS22, DBLP:conf/msr/CiniselliCPPPB21, guo2021learning} \\
\cline{2-5}
&Edit similarity& [S8]&3.81 &\cite{DBLP:conf/emnlp/ClementLLTDDSS21, DBLP:conf/sigsoft/SvyatkovskiyDFS20, DBLP:conf/acl/LuDHGHS22, DBLP:conf/icse/WenA0LB21} \\
\cline{2-5}
&Function similarity& [S9]&4.09 &\cite{guo2021learning} \\
\cline{2-5}
&Readability &[S10]&4.31 & - \\
\cline{2-5}
&Grammatical correctness &[S11]& 4.42& - \\
\cline{2-5}
&Others & & &\cite{ DBLP:conf/kbse/YangJ0SGL17, DBLP:conf/kbse/NguyenNLW19, DBLP:conf/sigsoft/SvyatkovskiyDFS20, DBLP:conf/emnlp/ClementLLTDDSS21, DBLP:conf/icse/WenA0LB21} \\\cline{2-5}
&\multicolumn{4}{|c|}{\cellcolor{lightgray}Access to service} \\
&Online & [S12]& 3.54&\cite{DBLP:conf/icse/WenA0LB21, DBLP:conf/sigsoft/SvyatkovskiyDFS20, DBLP:conf/icse/IzadiGG22} \\
\cline{2-5}
&Offline & [S14] & 3.89
&\cite{DBLP:conf/kbse/YangJ0SGL17} \\\cline{2-5}
&\multicolumn{4}{|c|}{\cellcolor{lightgray}Time Consumption} \\
&Mentioned & & &\cite{DBLP:conf/kbse/NguyenNLW19, DBLP:conf/emnlp/ClementLLTDDSS21, DBLP:conf/sigsoft/SvyatkovskiyDFS20, DBLP:conf/kbse/YangJ0SGL17} \\\hline
\end{tabular}
}
\end{table}

\subsection{RQ4: Current State-of-the-art Research}
After our literary review, we identify 26 papers in total from the top conference and journals in software engineering and artificial intelligence communities. Table \ref{tab:SOTA} shows the capabilities of surveyed code completion techniques. We can observe that the research on token-level completion is much more than that on statement-level completion (17 papers v.s. 10 papers).


\subsubsection{Token-level completion}

\noindent\textbf{Usage scenarios.} As seen from Table \ref{tab:SOTA}, all of the 17 collected token-level completion papers support
\textit{API recommendation}. However, only a few of these papers work on \textit{identifier completion}, and none of them mention \textit{path completion}. 
This may be attributed to that most identifiers and path tokens are Out-of-Vocabulary (OoV) words for the proposed techniques.

\noindent\textbf{Evaluation metrics.}
In this part, \textit{accuracy} is equivalent to top-1 accuracy and error rate, \textit{recall} refers to the more general top-k accuracy, and \textit{average rank} refers to the average rankings of the appropriate token in the prediction candidates (e.g., Mean Reciprocal Rank, MRR). In Table \ref{tab:SOTA}, we find that most of the papers utilize \textit{recall} and \textit{averaged rank} to evaluate the effectiveness of their approaches.
Moreover, other evaluation metrics are explored \cite{DBLP:conf/sigsoft/HellendoornD17, DBLP:conf/icse/HindleBSGD12, DBLP:conf/msr/SvyatkovskiyLHR21, DBLP:conf/kdd/SvyatkovskiyZFS19}. Some researchers attempt to explore better metrics as the proxy for users' productivity such as saved keystrokes \cite{DBLP:conf/icse/HindleBSGD12}. 


\noindent\textbf{Access to service.}
Few papers have focused on employing their techniques in industrial products. We classify the papers into two categories  including online and offline according to their deployment.
Svyatkovskiy et al. \cite{DBLP:conf/kdd/SvyatkovskiyZFS19, DBLP:conf/msr/SvyatkovskiyLHR21} develop their system as part of Intellicode extension in Visual Studio Code IDE \cite{vscode}, allowing programmers to use it offline. Izadi et al. \cite{DBLP:conf/icse/IzadiGG22} utilize a Transformer architecture, claiming that they put the model on the cloud and provide completion web services. Moreover, Hindle et al. \cite{DBLP:conf/icse/HindleBSGD12} mention that their tool is incorporated into the offline Eclipse plug-in.

\noindent\textbf{Efficiency.} Among collected token-level code completion papers, 11 of them explicitly consider the time consumption. Nine approaches proposed in \cite{DBLP:conf/kdd/SvyatkovskiyZFS19, DBLP:conf/msr/SvyatkovskiyLHR21, DBLP:conf/icse/IzadiGG22, DBLP:conf/icse/HindleBSGD12, DBLP:conf/sigsoft/HellendoornD17, DBLP:conf/sigsoft/TuSD14, DBLP:conf/iwpc/LiuLW0FJ20, DBLP:conf/oopsla/RaychevBV16, DBLP:conf/icml/BielikRV16} take less than 200 milliseconds to predict a token. 
According to our results, these techniques with such latency can satisfy 85\% practitioners. However, another two methods take several seconds to complete a single token \cite{DBLP:conf/sigsoft/NguyenNNN13, DBLP:conf/pldi/RaychevVY14}, which is considered unacceptable for token-level completion.

\subsubsection{Statement-level completion}

\noindent\textbf{Usage scenarios.} As seen from Table \ref{tab:SOTA}, most papers focus on \textit{next line prediction} which receives only 43\% support rate, but the more expected scenarios \textit{completion of currently edited line} and \textit{API argument recommendation} obtain less attention. Three papers propose approaches for \textit{block content prediction} scenario.
Moreover, no papers propose techniques to predict the methods' skeletons (\textit{skeleton prediction}) and complete string content (\textit{string completion}).

\finding{9}{Most papers focus on \textit{next line prediction} scenario, while only a few of them explore \textit{completion of currently edited line} and \textit{API argument recommendation}. Besides,
no paper has proposed techniques for \textit{skeleton prediction}, which is considered as the most expected scenario.}
\noindent\textbf{Evaluation metrics.}
From Table \ref{tab:SOTA}, most papers evaluate the generated code via \textit{overlapped n-grams} such as BLUE and ROUGE, and \textit{edit similarity}, which receive the least support rate (i.e., 65\% and 70\%, respectively) according to our survey.
Only one paper evaluates the generated code against human-written code via \textit{function/structure similarity}. Besides, none of them mention the \textit{readability} and \textit{grammatical correctness} of the generated code. Some papers also propose customized metrics to comprehensively evaluate their statement-level completion approaches such as click-through rate \cite{ DBLP:conf/kbse/YangJ0SGL17, DBLP:conf/kbse/NguyenNLW19, DBLP:conf/sigsoft/SvyatkovskiyDFS20, DBLP:conf/emnlp/ClementLLTDDSS21, DBLP:conf/icse/WenA0LB21}.
\finding{10}{Most papers focus on measuring \textit{overlapped n-grams} and \textit{edit similarity} between the generated code and the human-written code that are not preferred by the large majority of participants. No paper evaluates the generated code via \textit{grammatical correctness} and \textit{readability}, which the practitioners value most.}

\noindent\textbf{Access to service.}
The work \cite{DBLP:conf/sigsoft/SvyatkovskiyDFS20} proposes a tool called Intellicode Compose, which deploys the models on the cloud and also allow
client-side caching. Moreover, Wen et al. \cite{DBLP:conf/icse/WenA0LB21} present an android studio plugin as a web service. In addition, the tool proposed in \cite{DBLP:conf/kbse/YangJ0SGL17} serves as an Eclipse plug-in and provide offline completion service.

\noindent\textbf{Efficiency.}
Among the collected papers that propose statement-level code completion approaches, only four of them explicitly discuss the time consumption of their techniques. For example, the methods proposed in \cite{DBLP:conf/msr/CiniselliCPPPB21, DBLP:conf/kbse/YangJ0SGL17, DBLP:conf/sigsoft/SvyatkovskiyDFS20} take less than 1 second to complete code statements, while the work in \cite{DBLP:conf/kbse/NguyenNLW19} takes more than 5.5 seconds on average for prediction.

\finding{11}{ Time consumption, a critical adoption factor of code completion tools, is missing in 40\% of our collected token-level and statement-level papers.
}




\section{Discussion}\label{sec:discuss}

\subsection{Implications}
Our survey results highlight several implications
for the research community:
\subsubsection{Implication on code completion scenarios}
For token-level code completion tools, besides \textit{identifier completion} and \textit{API recommendation}, they are also expected to support \textit{path completion} which is expected by 86\% participants.

For the code completion tools that are capable of predicting lines of code, the most important scenarios for programmers are \textit{skeleton prediction}, \textit{completion of currently edited line} and \textit{API argument recommendation}. However, most current papers about statement-level completion focus on less anticipated scenarios such as \textit{next line prediction} \cite{guo2021learning, DBLP:conf/icse/IzadiGG22}.  

\subsubsection{Implication on code completion tools}
The large majority of programmers expect code completion tools to be more intelligent. For instance, 76\% participants expect
the tools to learn their programming habits so that they can keep the programming style consistent and reduce modification. 
Besides, most of them are also eager to be informed with additional information about the predicted code such as API definition or documentation (81\%). One participant stated that ``\textit{It is hard for me to remember countless APIs in my project. When I use an API, if the code completion tool locates and shows the API documentation simultaneously, it can significantly save time in searching the API.}"

Besides, most practitioners wish the time consumption for tool installation and configuration to be within
20 minutes. If they have to spend much time on tool installation, they may be dissuaded from using it.
\subsubsection{Implication on evaluation metric}
Evaluation metric is another important factor that should satisfy practitioners' expectations. Most of existing studies about statement-level code completion evaluate the generated code by comparing with human-written code in terms of \textit{overlapping n-grams} (such as BLEU score and ROUGE) \cite{DBLP:conf/msr/CiniselliCPPPB21} and \textit{editing similarity} \cite{DBLP:conf/nips/LuGRHSBCDJTLZSZ21, DBLP:conf/emnlp/ClementLLTDDSS21}. However, these two metrics receive the least support rate among all the evaluation metrics in our survey. Practitioners more expect the tools to use
the metric \textit{function/structure similarity} for evaluation. A participant told us: ``\textit{
... BLEU score may be a good criteria to judge the quality of generated natural language texts, however, it can hardly evaluate whether a code snippet is satisfactory."}
In addition, the metrics that participants value most such as \textit{grammatical correctness} and \textit{readability} of generated code are missing in current publications. 



\subsubsection{Balancing effectiveness and code quantity}
From our survey, 46\% practitioners regard the effectiveness of statement-level code completion weighs more than the quantity of predicted code (39\%). Considering that two completion strategies are both preferred by
a certain number of participants,
it will be better if the tools provide a configurable option for users to
decide the quantity of predicted code.

Besides, we also observe that most practitioners expect that the completed code does not need extensive manual modification.
For instance, 74\% participants cannot accept that more than 40\% of completed code needs manual modification. Thus, code completion tools may preliminarily estimate the probability of whether over 40\% of predicted code needs modifying. If the probability is high, tools can terminate current completion process and generate completion results.

\subsubsection{Implication on code completion latency}
Completion latency is one key aspect that substantially affects practitioners' likelihood to adopt code completion tools. From our survey, only 15\% participants can accept that tools take more than 400 milliseconds to predict a token. Besides, 83\% of them wish the average time consumption of generating a line of code to be less than 1 second, and 80\% participants expect the statement-level completion latency to be no more than 2 seconds. However, the time consumption factor is missing in nearly half of our collected papers. Researchers should pay more attention on code completion latency.
Considering that offline tools are the most expected, 
how to effectively compress the completion models is an important direction for future studies.
\subsubsection{Improving robustness of code completion tools} 
In our survey, many participants mention
that the robustness of code completion tools
is also vital to user experience. Robustness requires that the completion results are not affected by slight perturbations in the input. From our survey, a participant shared his/her experience with
Tabnine: ``\textit{Tabnine can successfully predict the API arguments if the variable is named `X\_train'. However, if I modify the name to `x\_train', the completion results will be totally different.}" Besides identifier changes, code completion tools are also supposed to be robust to statement changes. A participant stated that:
``\textit{AiXCoder can predict the block content of a method well, but the results turn to be terrible when I inserted an unrelated assignment statement.}''

\subsection{Threats to Validity}

One of the threat in our survey is that there may be some participants who do not fully understand the questions. For instance, some participants have never used statement-level code completion tools. Therefore, they may be unfamiliar with the questions of statement-level code completion. To reduce this threat, we utilize one or two clear images or GIFs to describe each scenario and facilitate them better understanding the questions. Furthermore, some participants do not answer the questions seriously and the results cannot reflect their beliefs. Therefore, we drop the responses completed less than two minutes. This is a common and tolerable threat to validity in previous studies, e.g., \cite{DBLP:journals/tse/KimZN14}.

Another threat is that our participants may not be representative in typical programmers. Our solution is to widely survey practitioners working in many IT companies. We believe we have made this threat have minimal impact on the results of our survey.
\section{Related Work}\label{sec:related}
\subsection{Code Completion}
Traditional code completion focused on static analysis techniques associated with manually defined rules to suggest code \cite{DBLP:conf/pldi/MandelinXBK05, DBLP:conf/pldi/GveroKKP13, DBLP:conf/pldi/PerelmanGBG12, DBLP:conf/kbse/ThummalapentaX07}. For instance, researchers utilized type information \cite{DBLP:conf/icse/HouP10, DBLP:conf/pldi/GveroKKP13}, similar code snippets \cite{DBLP:conf/sigsoft/BruchMM09} and history data \cite{DBLP:conf/kbse/RobbesL08} to predict needed code tokens.

Equipped with machine learning, a series of code completion work equipped with statistical language models was proposed \cite{DBLP:conf/icse/HindleBSGD12, DBLP:conf/icse/NguyenN15, DBLP:conf/se/ProkschLM16, DBLP:conf/sigsoft/HellendoornD17, DBLP:conf/sigsoft/NguyenNNN13, DBLP:conf/sigsoft/TuSD14}. 
For instance, Hellen et al. \cite{DBLP:conf/sigsoft/HellendoornD17} explicitly took the techniques such as nested scopes into account and improved the performance of the n-gram model. Moreover, Raychev et al. combined decision trees and domain knowledge, proposing a probabilistic model \cite{DBLP:conf/oopsla/RaychevBV16}. 

With the development of deep learning, neural networks such as RNN \cite{DBLP:journals/corr/abs-1808-03314} and Transformer \cite{DBLP:conf/nips/VaswaniSPUJGKP17} showed great capability to learn features from source code \cite{DBLP:conf/sigsoft/HellendoornD17, DBLP:conf/icse/KimZT021, DBLP:conf/ijcai/LiWLK18}. For instance, Li et al. \cite{DBLP:conf/ijcai/LiWLK18} proposed a point mixture network to relieve the Out-of-Vocabulary problem. Kim et al. \cite{DBLP:conf/icse/KimZT021} parsed the code into abstract syntax trees and fed the trees into a Transformer-based model. 
In recent years, pre-trained language models have been leveraged for predicting multiple code tokens \cite{DBLP:conf/sigsoft/SvyatkovskiyDFS20, DBLP:conf/icse/IzadiGG22}. Svyatkovskiy et al. \cite{DBLP:conf/sigsoft/SvyatkovskiyDFS20} proposed GPT-C, which could predict code statements. 
Izadi et al. \cite{DBLP:conf/icse/IzadiGG22} raised CodeFill, which combined type and semantic information of code, further improving the completion performance.

\subsection{Studies on Code Completion Practices}
Apart from research on code completion, some other work focus on studying code completion practices \cite{DBLP:conf/chi/Vaithilingam0G22, DBLP:conf/pldi/0001KLRRSSA22, DBLP:conf/icse/HellendoornPGB19, DBLP:conf/icse/AyeKL21}. For instance, Vaithilingam et al. \cite{DBLP:conf/chi/Vaithilingam0G22} asked participants to finish programming tasks with or without Copilot \cite{copilot} and determined whether Copilot is useful. In addition, Ziegler et al. \cite{DBLP:conf/pldi/0001KLRRSSA22} focused on investigating evaluation metrics of code completion. Authors compared the measurable user data (objective data) and the user-reported productivity (subjective data), and identified the most representative metric as a proxy of productivity. The work \cite{DBLP:conf/icse/HellendoornPGB19, DBLP:conf/icse/AyeKL21} both pointed out the difference between synthetic data and real-world data, and identified the most important tokens that needed to be completed.

However, no prior studies have investigated the practitioners' expectations on code completion. In this paper, we conduct a large scale user study, investigating practitioners' expectations on multiple aspects of code completion tools. Moreover, we perform a comprehensive literature review to reveal the gap between current techniques and practitioners' expectations, providing future directions for researchers.
\section{Conclusion and Future Work}
In this paper, we interview 15 professionals and survey 599 practitioners on completion practices, issues they face and their expectations on code completion tools. Practitioners expect code completion tools to suggest code for different granularities. Practitioners also expect a code completion tool to satisfy the aspects including \textit{usage scenarios}, \textit{evaluation metrics}, \textit{access to service}, \textit{effectiveness} and \textit{efficiency}. We also compare the capability of current research with practitioners' expectations via a literature review, pointing out the aspects to be improved for meeting the demands of practitioners.

\bibliographystyle{IEEEtran}
\bibliography{sample}

\begin{thebibliography}{10}
\providecommand{\url}[1]{#1}
\csname url@samestyle\endcsname
\providecommand{\newblock}{\relax}
\providecommand{\bibinfo}[2]{#2}
\providecommand{\BIBentrySTDinterwordspacing}{\spaceskip=0pt\relax}
\providecommand{\BIBentryALTinterwordstretchfactor}{4}
\providecommand{\BIBentryALTinterwordspacing}{\spaceskip=\fontdimen2\font plus
\BIBentryALTinterwordstretchfactor\fontdimen3\font minus
  \fontdimen4\font\relax}
\providecommand{\BIBforeignlanguage}[2]{{%
\expandafter\ifx\csname l@#1\endcsname\relax
\typeout{** WARNING: IEEEtran.bst: No hyphenation pattern has been}%
\typeout{** loaded for the language `#1'. Using the pattern for}%
\typeout{** the default language instead.}%
\else
\language=\csname l@#1\endcsname
\fi
#2}}
\providecommand{\BIBdecl}{\relax}
\BIBdecl

\bibitem{DBLP:conf/icsm/LiHLYT21}
J.~Li, R.~Huang, W.~Li, K.~Yao, and W.~Tan, ``Toward less hidden cost of code
  completion with acceptance and ranking models,'' in \emph{{IEEE}
  International Conference on Software Maintenance and Evolution, {ICSME} 2021,
  Luxembourg, September 27 - October 1, 2021}.\hskip 1em plus 0.5em minus
  0.4em\relax {IEEE}, 2021, pp. 195--205.

\bibitem{DBLP:conf/wcre/AmannPNM16}
S.~Amann, S.~Proksch, S.~Nadi, and M.~Mezini, ``A study of visual studio usage
  in practice,'' in \emph{{IEEE} 23rd International Conference on Software
  Analysis, Evolution, and Reengineering, {SANER} 2016, Suita, Osaka, Japan,
  March 14-18, 2016 - Volume 1}.\hskip 1em plus 0.5em minus 0.4em\relax {IEEE}
  Computer Society, 2016, pp. 124--134.

\bibitem{DBLP:conf/kdd/SvyatkovskiyZFS19}
A.~Svyatkovskiy, Y.~Zhao, S.~Fu, and N.~Sundaresan, ``Pythia: Ai-assisted code
  completion system,'' in \emph{Proceedings of the 25th {ACM} {SIGKDD}
  International Conference on Knowledge Discovery {\&} Data Mining, {KDD} 2019,
  Anchorage, AK, USA, August 4-8, 2019}, A.~Teredesai, V.~Kumar, Y.~Li,
  R.~Rosales, E.~Terzi, and G.~Karypis, Eds.\hskip 1em plus 0.5em minus
  0.4em\relax {ACM}, 2019, pp. 2727--2735.

\bibitem{DBLP:conf/aaai/WangL21a}
Y.~Wang and H.~Li, ``Code completion by modeling flattened abstract syntax
  trees as graphs,'' in \emph{Thirty-Fifth {AAAI} Conference on Artificial
  Intelligence, {AAAI} 2021, Thirty-Third Conference on Innovative Applications
  of Artificial Intelligence, {IAAI} 2021, The Eleventh Symposium on
  Educational Advances in Artificial Intelligence, {EAAI} 2021, Virtual Event,
  February 2-9, 2021}.\hskip 1em plus 0.5em minus 0.4em\relax {AAAI} Press,
  2021, pp. 14\,015--14\,023.

\bibitem{DBLP:conf/icse/KimZT021}
S.~Kim, J.~Zhao, Y.~Tian, and S.~Chandra, ``Code prediction by feeding trees to
  transformers,'' in \emph{43rd {IEEE/ACM} International Conference on Software
  Engineering, {ICSE} 2021, Madrid, Spain, 22-30 May 2021}.\hskip 1em plus
  0.5em minus 0.4em\relax {IEEE}, 2021, pp. 150--162.

\bibitem{DBLP:conf/iwpc/LiuLW0FJ20}
F.~Liu, G.~Li, B.~Wei, X.~Xia, Z.~Fu, and Z.~Jin, ``A self-attentional neural
  architecture for code completion with multi-task learning,'' in \emph{{ICPC}
  '20: 28th International Conference on Program Comprehension, Seoul, Republic
  of Korea, July 13-15, 2020}.\hskip 1em plus 0.5em minus 0.4em\relax {ACM},
  2020, pp. 37--47.

\bibitem{DBLP:conf/msr/SvyatkovskiyLHR21}
A.~Svyatkovskiy, S.~Lee, A.~Hadjitofi, M.~Riechert, J.~V. Franco, and
  M.~Allamanis, ``Fast and memory-efficient neural code completion,'' in
  \emph{18th {IEEE/ACM} International Conference on Mining Software
  Repositories, {MSR} 2021, Madrid, Spain, May 17-19, 2021}.\hskip 1em plus
  0.5em minus 0.4em\relax {IEEE}, 2021, pp. 329--340.

\bibitem{DBLP:conf/sigsoft/SvyatkovskiyDFS20}
A.~Svyatkovskiy, S.~K. Deng, S.~Fu, and N.~Sundaresan, ``Intellicode compose:
  code generation using transformer,'' in \emph{{ESEC/FSE} '20: 28th {ACM}
  Joint European Software Engineering Conference and Symposium on the
  Foundations of Software Engineering, Virtual Event, USA, November 8-13,
  2020}, P.~Devanbu, M.~B. Cohen, and T.~Zimmermann, Eds.\hskip 1em plus 0.5em
  minus 0.4em\relax {ACM}, 2020, pp. 1433--1443.

\bibitem{guo2021learning}
D.~Guo, A.~Svyatkovskiy, J.~Yin, N.~Duan, M.~Brockschmidt, and M.~Allamanis,
  ``Learning to complete code with sketches,'' in \emph{International
  Conference on Learning Representations}, 2021.

\bibitem{DBLP:conf/icse/IzadiGG22}
M.~Izadi, R.~Gismondi, and G.~Gousios, ``Codefill: Multi-token code completion
  by jointly learning from structure and naming sequences,'' in \emph{44th
  {IEEE/ACM} 44th International Conference on Software Engineering, {ICSE}
  2022, Pittsburgh, PA, USA, May 25-27, 2022}.\hskip 1em plus 0.5em minus
  0.4em\relax {ACM}, 2022, pp. 401--412.

\bibitem{copilot}
``Copilot,'' \url{https://github.com/features/copilot/}.

\bibitem{aixcoder}
``Aixcoder,'' \url{https://aixcoder.com/}.

\bibitem{tabnine}
``Tabnine,'' \url{https://www.tabnine.com/}.

\bibitem{clangd}
``Clangd,'' \url{https://clangd.llvm.org/}.

\bibitem{DBLP:conf/pldi/MandelinXBK05}
D.~Mandelin, L.~Xu, R.~Bod{\'{\i}}k, and D.~Kimelman, ``Jungloid mining:
  helping to navigate the {API} jungle,'' in \emph{Proceedings of the {ACM}
  {SIGPLAN} 2005 Conference on Programming Language Design and Implementation,
  Chicago, IL, USA, June 12-15, 2005}, V.~Sarkar and M.~W. Hall, Eds.\hskip 1em
  plus 0.5em minus 0.4em\relax {ACM}, 2005, pp. 48--61.

\bibitem{DBLP:conf/pldi/GveroKKP13}
T.~Gvero, V.~Kuncak, I.~Kuraj, and R.~Piskac, ``Complete completion using types
  and weights,'' in \emph{{ACM} {SIGPLAN} Conference on Programming Language
  Design and Implementation, {PLDI} '13, Seattle, WA, USA, June 16-19, 2013},
  H.~Boehm and C.~Flanagan, Eds.\hskip 1em plus 0.5em minus 0.4em\relax {ACM},
  2013, pp. 27--38.

\bibitem{DBLP:conf/icse/HindleBSGD12}
A.~Hindle, E.~T. Barr, Z.~Su, M.~Gabel, and P.~T. Devanbu, ``On the naturalness
  of software,'' in \emph{34th International Conference on Software
  Engineering, {ICSE} 2012, June 2-9, 2012, Zurich, Switzerland}, M.~Glinz,
  G.~C. Murphy, and M.~Pezz{\`{e}}, Eds.\hskip 1em plus 0.5em minus 0.4em\relax
  {IEEE} Computer Society, 2012, pp. 837--847.

\bibitem{DBLP:conf/se/ProkschLM16}
S.~Proksch, J.~Lerch, and M.~Mezini, ``Intelligent code completion with
  bayesian networks,'' in \emph{Software Engineering 2016, Fachtagung des
  GI-Fachbereichs Softwaretechnik, 23.-26. Februar 2016, Wien,
  {\"{O}}sterreich}, ser. {LNI}, J.~Knoop and U.~Zdun, Eds., vol.
  {P-252}.\hskip 1em plus 0.5em minus 0.4em\relax {GI}, 2016, pp. 25--26.

\bibitem{DBLP:conf/icse/NguyenN15}
A.~T. Nguyen and T.~N. Nguyen, ``Graph-based statistical language model for
  code,'' in \emph{37th {IEEE/ACM} International Conference on Software
  Engineering, {ICSE} 2015, Florence, Italy, May 16-24, 2015, Volume 1},
  A.~Bertolino, G.~Canfora, and S.~G. Elbaum, Eds.\hskip 1em plus 0.5em minus
  0.4em\relax {IEEE} Computer Society, 2015, pp. 858--868.

\bibitem{DBLP:conf/kbse/NguyenNLW19}
S.~V. Nguyen, T.~N. Nguyen, Y.~Li, and S.~Wang, ``Combining program analysis
  and statistical language model for code statement completion,'' in \emph{34th
  {IEEE/ACM} International Conference on Automated Software Engineering, {ASE}
  2019, San Diego, CA, USA, November 11-15, 2019}.\hskip 1em plus 0.5em minus
  0.4em\relax {IEEE}, 2019, pp. 710--721.

\bibitem{DBLP:conf/msr/CiniselliCPPPB21}
M.~Ciniselli, N.~Cooper, L.~Pascarella, D.~Poshyvanyk, M.~D. Penta, and
  G.~Bavota, ``An empirical study on the usage of {BERT} models for code
  completion,'' in \emph{18th {IEEE/ACM} International Conference on Mining
  Software Repositories, {MSR} 2021, Madrid, Spain, May 17-19, 2021}.\hskip 1em
  plus 0.5em minus 0.4em\relax {IEEE}, 2021, pp. 108--119.

\bibitem{DBLP:conf/naacl/DevlinCLT19}
J.~Devlin, M.~Chang, K.~Lee, and K.~Toutanova, ``{BERT:} pre-training of deep
  bidirectional transformers for language understanding,'' in \emph{Proceedings
  of the 2019 Conference of the North American Chapter of the Association for
  Computational Linguistics: Human Language Technologies, {NAACL-HLT} 2019,
  Minneapolis, MN, USA, June 2-7, 2019, Volume 1 (Long and Short Papers)},
  J.~Burstein, C.~Doran, and T.~Solorio, Eds.\hskip 1em plus 0.5em minus
  0.4em\relax Association for Computational Linguistics, 2019, pp. 4171--4186.

\bibitem{brown2020language}
T.~Brown, B.~Mann, N.~Ryder, M.~Subbiah, J.~D. Kaplan, P.~Dhariwal,
  A.~Neelakantan, P.~Shyam, G.~Sastry, A.~Askell \emph{et~al.}, ``Language
  models are few-shot learners,'' \emph{Advances in neural information
  processing systems}, vol.~33, pp. 1877--1901, 2020.

\bibitem{radford2019language}
A.~Radford, J.~Wu, R.~Child, D.~Luan, D.~Amodei, I.~Sutskever \emph{et~al.},
  ``Language models are unsupervised multitask learners,'' \emph{OpenAI blog},
  vol.~1, no.~8, p.~9, 2019.

\bibitem{DBLP:conf/icse/HuX0WCZ22}
X.~Hu, X.~Xia, D.~Lo, Z.~Wan, Q.~Chen, and T.~Zimmermann, ``Practitioners'
  expectations on automated code comment generation,'' in \emph{44th {IEEE/ACM}
  44th International Conference on Software Engineering, {ICSE} 2022,
  Pittsburgh, PA, USA, May 25-27, 2022}.\hskip 1em plus 0.5em minus 0.4em\relax
  {ACM}, 2022, pp. 1693--1705.

\bibitem{DBLP:conf/issta/KochharXLL16}
P.~S. Kochhar, X.~Xia, D.~Lo, and S.~Li, ``Practitioners' expectations on
  automated fault localization,'' in \emph{Proceedings of the 25th
  International Symposium on Software Testing and Analysis, {ISSTA} 2016,
  Saarbr{\"{u}}cken, Germany, July 18-20, 2016}, A.~Zeller and A.~Roychoudhury,
  Eds.\hskip 1em plus 0.5em minus 0.4em\relax {ACM}, 2016, pp. 165--176.

\bibitem{wj}
``Tencent questionnair,'' \url{https://wj.qq.com/}.

\bibitem{wj2}
``Google forms,'' \url{https://docs.google.com/forms}.

\bibitem{DBLP:conf/nips/LuGRHSBCDJTLZSZ21}
S.~Lu, D.~Guo, S.~Ren, J.~Huang, A.~Svyatkovskiy, A.~Blanco, C.~B. Clement,
  D.~Drain, D.~Jiang, D.~Tang, G.~Li, L.~Zhou, L.~Shou, L.~Zhou, M.~Tufano,
  M.~Gong, M.~Zhou, N.~Duan, N.~Sundaresan, S.~K. Deng, S.~Fu, and S.~Liu,
  ``Codexglue: {A} machine learning benchmark dataset for code understanding
  and generation,'' in \emph{Proceedings of the Neural Information Processing
  Systems Track on Datasets and Benchmarks 1, NeurIPS Datasets and Benchmarks
  2021, December 2021, virtual}, J.~Vanschoren and S.~Yeung, Eds., 2021.

\bibitem{DBLP:conf/acl/PapineniRWZ02}
K.~Papineni, S.~Roukos, T.~Ward, and W.~Zhu, ``Bleu: a method for automatic
  evaluation of machine translation,'' in \emph{Proceedings of the 40th Annual
  Meeting of the Association for Computational Linguistics, July 6-12, 2002,
  Philadelphia, PA, {USA}}.\hskip 1em plus 0.5em minus 0.4em\relax {ACL}, 2002,
  pp. 311--318.

\bibitem{DBLP:conf/ijcai/LiWLK18}
J.~Li, Y.~Wang, M.~R. Lyu, and I.~King, ``Code completion with neural attention
  and pointer networks,'' in \emph{Proceedings of the Twenty-Seventh
  International Joint Conference on Artificial Intelligence, {IJCAI} 2018, July
  13-19, 2018, Stockholm, Sweden}, J.~Lang, Ed.\hskip 1em plus 0.5em minus
  0.4em\relax ijcai.org, 2018, pp. 4159--4165.

\bibitem{DBLP:conf/icml/BielikRV16}
P.~Bielik, V.~Raychev, and M.~T. Vechev, ``{PHOG:} probabilistic model for
  code,'' in \emph{Proceedings of the 33nd International Conference on Machine
  Learning, {ICML} 2016, New York City, NY, USA, June 19-24, 2016}, ser. {JMLR}
  Workshop and Conference Proceedings, M.~Balcan and K.~Q. Weinberger, Eds.,
  vol.~48.\hskip 1em plus 0.5em minus 0.4em\relax JMLR.org, 2016, pp.
  2933--2942.

\bibitem{DBLP:conf/oopsla/RaychevBV16}
V.~Raychev, P.~Bielik, and M.~T. Vechev, ``Probabilistic model for code with
  decision trees,'' in \emph{Proceedings of the 2016 {ACM} {SIGPLAN}
  International Conference on Object-Oriented Programming, Systems, Languages,
  and Applications, {OOPSLA} 2016, part of {SPLASH} 2016, Amsterdam, The
  Netherlands, October 30 - November 4, 2016}, E.~Visser and Y.~Smaragdakis,
  Eds.\hskip 1em plus 0.5em minus 0.4em\relax {ACM}, 2016, pp. 731--747.

\bibitem{DBLP:conf/sigsoft/NguyenNNN13}
T.~T. Nguyen, A.~T. Nguyen, H.~A. Nguyen, and T.~N. Nguyen, ``A statistical
  semantic language model for source code,'' in \emph{Joint Meeting of the
  European Software Engineering Conference and the {ACM} {SIGSOFT} Symposium on
  the Foundations of Software Engineering, ESEC/FSE'13, Saint Petersburg,
  Russian Federation, August 18-26, 2013}, B.~Meyer, L.~Baresi, and M.~Mezini,
  Eds.\hskip 1em plus 0.5em minus 0.4em\relax {ACM}, 2013, pp. 532--542.

\bibitem{DBLP:conf/kbse/LiuLZJ20}
F.~Liu, G.~Li, Y.~Zhao, and Z.~Jin, ``Multi-task learning based pre-trained
  language model for code completion,'' in \emph{35th {IEEE/ACM} International
  Conference on Automated Software Engineering, {ASE} 2020, Melbourne,
  Australia, September 21-25, 2020}.\hskip 1em plus 0.5em minus 0.4em\relax
  {IEEE}, 2020, pp. 473--485.

\bibitem{DBLP:conf/sigsoft/HellendoornD17}
V.~J. Hellendoorn and P.~T. Devanbu, ``Are deep neural networks the best choice
  for modeling source code?'' in \emph{Proceedings of the 2017 11th Joint
  Meeting on Foundations of Software Engineering, {ESEC/FSE} 2017, Paderborn,
  Germany, September 4-8, 2017}, E.~Bodden, W.~Sch{\"{a}}fer, A.~van Deursen,
  and A.~Zisman, Eds.\hskip 1em plus 0.5em minus 0.4em\relax {ACM}, 2017, pp.
  763--773.

\bibitem{DBLP:conf/acl/LuDHGHS22}
S.~Lu, N.~Duan, H.~Han, D.~Guo, S.~Hwang, and A.~Svyatkovskiy, ``Reacc: {A}
  retrieval-augmented code completion framework,'' in \emph{Proceedings of the
  60th Annual Meeting of the Association for Computational Linguistics (Volume
  1: Long Papers), {ACL} 2022, Dublin, Ireland, May 22-27, 2022}, S.~Muresan,
  P.~Nakov, and A.~Villavicencio, Eds.\hskip 1em plus 0.5em minus 0.4em\relax
  Association for Computational Linguistics, 2022, pp. 6227--6240.

\bibitem{DBLP:conf/sigsoft/TuSD14}
Z.~Tu, Z.~Su, and P.~T. Devanbu, ``On the localness of software,'' in
  \emph{Proceedings of the 22nd {ACM} {SIGSOFT} International Symposium on
  Foundations of Software Engineering, (FSE-22), Hong Kong, China, November 16
  - 22, 2014}, S.~Cheung, A.~Orso, and M.~D. Storey, Eds.\hskip 1em plus 0.5em
  minus 0.4em\relax {ACM}, 2014, pp. 269--280.

\bibitem{DBLP:conf/pldi/RaychevVY14}
V.~Raychev, M.~T. Vechev, and E.~Yahav, ``Code completion with statistical
  language models,'' in \emph{{ACM} {SIGPLAN} Conference on Programming
  Language Design and Implementation, {PLDI} '14, Edinburgh, United Kingdom -
  June 09 - 11, 2014}, M.~F.~P. O'Boyle and K.~Pingali, Eds.\hskip 1em plus
  0.5em minus 0.4em\relax {ACM}, 2014, pp. 419--428.

\bibitem{DBLP:conf/emnlp/ClementLLTDDSS21}
C.~B. Clement, S.~Lu, X.~Liu, M.~Tufano, D.~Drain, N.~Duan, N.~Sundaresan, and
  A.~Svyatkovskiy, ``Long-range modeling of source code files with ewash:
  Extended window access by syntax hierarchy,'' in \emph{Proceedings of the
  2021 Conference on Empirical Methods in Natural Language Processing, {EMNLP}
  2021, Virtual Event / Punta Cana, Dominican Republic, 7-11 November, 2021},
  M.~Moens, X.~Huang, L.~Specia, and S.~W. Yih, Eds.\hskip 1em plus 0.5em minus
  0.4em\relax Association for Computational Linguistics, 2021, pp. 4713--4722.

\bibitem{DBLP:conf/kbse/YangJ0SGL17}
Y.~Yang, Y.~Jiang, M.~Gu, J.~Sun, J.~Gao, and H.~Liu, ``A language model for
  statements of software code,'' in \emph{Proceedings of the 32nd {IEEE/ACM}
  International Conference on Automated Software Engineering, {ASE} 2017,
  Urbana, IL, USA, October 30 - November 03, 2017}, G.~Rosu, M.~D. Penta, and
  T.~N. Nguyen, Eds.\hskip 1em plus 0.5em minus 0.4em\relax {IEEE} Computer
  Society, 2017, pp. 682--687.

\bibitem{DBLP:conf/icse/WenA0LB21}
F.~Wen, E.~Aghajani, C.~Nagy, M.~Lanza, and G.~Bavota, ``Siri, write the next
  method,'' in \emph{43rd {IEEE/ACM} International Conference on Software
  Engineering, {ICSE} 2021, Madrid, Spain, 22-30 May 2021}.\hskip 1em plus
  0.5em minus 0.4em\relax {IEEE}, 2021, pp. 138--149.

\bibitem{vscode}
``Visual studio code,'' \url{https://code.visualstudio.com/}.

\bibitem{DBLP:journals/tse/KimZN14}
M.~Kim, T.~Zimmermann, and N.~Nagappan, ``An empirical study of
  refactoringchallenges and benefits at microsoft,'' \emph{{IEEE} Trans.
  Software Eng.}, vol.~40, no.~7, pp. 633--649, 2014.

\bibitem{DBLP:conf/pldi/PerelmanGBG12}
D.~Perelman, S.~Gulwani, T.~Ball, and D.~Grossman, ``Type-directed completion
  of partial expressions,'' in \emph{{ACM} {SIGPLAN} Conference on Programming
  Language Design and Implementation, {PLDI} '12, Beijing, China - June 11 -
  16, 2012}, J.~Vitek, H.~Lin, and F.~Tip, Eds.\hskip 1em plus 0.5em minus
  0.4em\relax {ACM}, 2012, pp. 275--286.

\bibitem{DBLP:conf/kbse/ThummalapentaX07}
S.~Thummalapenta and T.~Xie, ``Parseweb: a programmer assistant for reusing
  open source code on the web,'' in \emph{22nd {IEEE/ACM} International
  Conference on Automated Software Engineering {(ASE} 2007), November 5-9,
  2007, Atlanta, Georgia, {USA}}, R.~E.~K. Stirewalt, A.~Egyed, and B.~Fischer,
  Eds.\hskip 1em plus 0.5em minus 0.4em\relax {ACM}, 2007, pp. 204--213.

\bibitem{DBLP:conf/icse/HouP10}
D.~Hou and D.~M. Pletcher, ``Towards a better code completion system by {API}
  grouping, filtering, and popularity-based ranking,'' in \emph{Proceedings of
  the 2nd International Workshop on Recommendation Systems for Software
  Engineering, {RSSE} 2010, Cape Town, South Africa, May 4, 2010}, R.~Holmes,
  M.~P. Robillard, R.~J. Walker, and T.~Zimmermann, Eds.\hskip 1em plus 0.5em
  minus 0.4em\relax {ACM}, 2010, pp. 26--30.

\bibitem{DBLP:conf/sigsoft/BruchMM09}
M.~Bruch, M.~Monperrus, and M.~Mezini, ``Learning from examples to improve code
  completion systems,'' in \emph{Proceedings of the 7th joint meeting of the
  European Software Engineering Conference and the {ACM} {SIGSOFT}
  International Symposium on Foundations of Software Engineering, 2009,
  Amsterdam, The Netherlands, August 24-28, 2009}, H.~van Vliet and V.~Issarny,
  Eds.\hskip 1em plus 0.5em minus 0.4em\relax {ACM}, 2009, pp. 213--222.

\bibitem{DBLP:conf/kbse/RobbesL08}
R.~Robbes and M.~Lanza, ``How program history can improve code completion,'' in
  \emph{23rd {IEEE/ACM} International Conference on Automated Software
  Engineering {(ASE} 2008), 15-19 September 2008, L'Aquila, Italy}.\hskip 1em
  plus 0.5em minus 0.4em\relax {IEEE} Computer Society, 2008, pp. 317--326.

\bibitem{DBLP:journals/corr/abs-1808-03314}
A.~Sherstinsky, ``Fundamentals of recurrent neural network {(RNN)} and long
  short-term memory {(LSTM)} network,'' \emph{CoRR}, vol. abs/1808.03314, 2018.

\bibitem{DBLP:conf/nips/VaswaniSPUJGKP17}
A.~Vaswani, N.~Shazeer, N.~Parmar, J.~Uszkoreit, L.~Jones, A.~N. Gomez,
  L.~Kaiser, and I.~Polosukhin, ``Attention is all you need,'' in
  \emph{Advances in Neural Information Processing Systems 30: Annual Conference
  on Neural Information Processing Systems 2017, December 4-9, 2017, Long
  Beach, CA, {USA}}, I.~Guyon, U.~von Luxburg, S.~Bengio, H.~M. Wallach,
  R.~Fergus, S.~V.~N. Vishwanathan, and R.~Garnett, Eds., 2017, pp. 5998--6008.

\bibitem{DBLP:conf/chi/Vaithilingam0G22}
P.~Vaithilingam, T.~Zhang, and E.~L. Glassman, ``Expectation vs. experience:
  Evaluating the usability of code generation tools powered by large language
  models,'' in \emph{{CHI} '22: {CHI} Conference on Human Factors in Computing
  Systems, New Orleans, LA, USA, 29 April 2022 - 5 May 2022, Extended
  Abstracts}, S.~D.~J. Barbosa, C.~Lampe, C.~Appert, and D.~A. Shamma,
  Eds.\hskip 1em plus 0.5em minus 0.4em\relax {ACM}, 2022, pp. 332:1--332:7.

\bibitem{DBLP:conf/pldi/0001KLRRSSA22}
A.~Ziegler, E.~Kalliamvakou, X.~A. Li, A.~Rice, D.~Rifkin, S.~Simister,
  G.~Sittampalam, and E.~Aftandilian, ``Productivity assessment of neural code
  completion,'' in \emph{MAPS@PLDI 2022: 6th {ACM} {SIGPLAN} International
  Symposium on Machine Programming, San Diego, CA, USA, 13 June 2022},
  S.~Chaudhuri and C.~Sutton, Eds.\hskip 1em plus 0.5em minus 0.4em\relax
  {ACM}, 2022, pp. 21--29.

\bibitem{DBLP:conf/icse/HellendoornPGB19}
V.~J. Hellendoorn, S.~Proksch, H.~C. Gall, and A.~Bacchelli, ``When code
  completion fails: a case study on real-world completions,'' in
  \emph{Proceedings of the 41st International Conference on Software
  Engineering, {ICSE} 2019, Montreal, QC, Canada, May 25-31, 2019}, J.~M.
  Atlee, T.~Bultan, and J.~Whittle, Eds.\hskip 1em plus 0.5em minus 0.4em\relax
  {IEEE} / {ACM}, 2019, pp. 960--970.

\bibitem{DBLP:conf/icse/AyeKL21}
G.~A. Aye, S.~Kim, and H.~Li, ``Learning autocompletion from real-world
  datasets,'' in \emph{43rd {IEEE/ACM} International Conference on Software
  Engineering: Software Engineering in Practice, {ICSE} {(SEIP)} 2021, Madrid,
  Spain, May 25-28, 2021}.\hskip 1em plus 0.5em minus 0.4em\relax {IEEE}, 2021,
  pp. 131--139.

\end{thebibliography}

\end{document}